# A decision support tool for ship biofouling management in the Baltic Sea

*Emilia Luoma[1,4,5], Mirka Laurila-Pant[1,4,5], Elias Altarriba[2,4], Inari Helle[3,4,5,6], Lena Granhag[7], Maiju Lehtiniemi [8], Greta Srėbalienė[9], Sergej Olenin[9], Annukka Lehikoinen[1,4,5]

[1] Ecosystems and Environment Research Programme, Faculty of Biological and Environmental Sciences, University of Helsinki, Finland

[2] South-Eastern Finland University of Applied Sciences (Xamk), Logistics and Seafaring, Kotka, Finland

[3] Organismal and Evolutionary Biology Research Programme, Faculty of Biological and Environmental Sciences, University of Helsinki, Finland

[4] Kotka Maritime Research Centre, Kotka, Finland

[5] Helsinki Institute of Sustainability Science (HELSUS), University of Helsinki, Finland

[6] Natural Resources Institute Finland (Luke), Helsinki, Finland

[7] Department of Mechanics and Maritime Sciences, Chalmers University of Technology, Gothenburg, Sweden

[8] Finnish Environment Institute, Marine Research Center, Helsinki, Finland

[9] Klaipeda University, Marine Research Institute, Klaipeda, Lithuania

## Abstract

Biofouling of ships causes major environmental and economic consequences all over the world. In addition, biofouling management of ship hulls causes both social, environmental and economic risks that should all be considered reaching well-balanced decisions. In addition, each case is unique and thus optimal management strategy must be considered case-specifically. We produced a novel decision support tool using Bayesian networks to promote the comprehensive understanding about the complex biofouling management issue in the Baltic Sea and to identify potential management options and their consequences. The tool compares the biofouling management strategies in relation to NIS (non-indigenous species) introduction risk, eco-toxicological risk due to biocidal coating, carbon dioxide emissions resulting from fuel consumption and costs related to fuel consumption, in-water cleaning and coating. According to the results, the optimal biofouling management strategy would consist of a biocidal-free coating with regular in-water cleaning and with devices collecting the material. However, the best biocidal-free coating type and the optimal in-water cleaning interval varies and depends e.g. on the operational profile of the ship. The decision support tool can increase the multi-perspective understanding about the issue and support the implementation of the optimal biofouling management strategies in the Baltic Sea.

## 1. Introduction

The International Maritime Organization (IMO) recognizes biofouling, the accumulation of organisms on surfaces such as the ship hull (Amara et al., 2018), as one of the main concerns of shipping. Biofouling causes both economic (Pagoropoulos et al., 2017) and environmental (Ojaveer et al., 2017) risks: increased fuel consumption and emissions, and a pathway for the spreading of non-indigenous species (NIS). NIS introductions create a major threat to marine ecosystems all over the world (Molnar et al., 2008). Shipping companies use biofouling management methods, coatings and in-water cleanings, to decrease the negative impacts of biofouling such as increased fuel costs (Schultz, 2007). However, the biocidal coatings diminishing the biofouling cause eco-toxicological load in the marine environment (Lagerström et al., 2018; Ytreberg et al., 2017). Further, the application of non-biocidal management options such as in-water cleaning (IWC) or biocidal free coatings is limited through environmental circumstances (e.g. partial ice covering hindering the use of fouling release coating) or legislation (Scianni & Georgiades, 2019). In the Baltic Sea, its special characteristics, low salinity, partial ice cover during winter and dense ship traffic, create extra challenges and thus biofouling management methods suitable for other marine environments cannot be transcribed directly to the Baltic Sea (Leppäranta & Myrberg, 2009).

Although the risks considering biofouling are well acknowledged, international legislation concerning ships' biofouling management is still missing (IMO, 2011). There is an urgent need to support the shipowners in 1) choosing the optimal biofouling management strategy, 2) preventing further introductions of NIS, and 3) reducing toxic load to the marine environment (Ojaveer & Kotta, 2015). All the various relevant economic, environmental and social aspects need to be considered case-specifically to reach well-balanced biofouling management. Methods enabling a holistic assessment of the system can support efficient management decisions and even the future enactment of legislation (HELCOM, 2010).

To meet this demand, we apply Bayesian Networks (BNs, e.g. Jensen & Nielsen, 2007) as a method to integrate available and applicable data and other knowledge to develop a decision analysis model for a probabilistic comparison of alternative biofouling management strategies in the Baltic Sea. The model serves as an interactive decision support tool (DST) that can be used for ship- and route-specific multi-objective comparisons. Further, it provides us with more general knowledge concerning the positive and negative aspects of divergent management strategies under various conditions. This approach can increase the systemic understanding concerning the complex biofouling management issue and the alternative perspectives to an optimal solution.

The tool developed in this paper is the first probabilistic quantitative tool considering several environmental and economic aspects: NIS introduction risk, eco-toxicological risk originating from biocidal coating, carbon dioxide emissions resulting from the fuel consumption of the ship (linked to level of biofouling), and costs related to fuel consumption and biofouling management, IWC and coating.

## 2. Materials and methods

### 2.1 Study area

Our study area is the Baltic Sea which is a unique brackish water area located in northern Europe. The special characteristics of the Baltic Sea such as shallowness, low salinity and partial ice cover during winter together with intensive marine traffic make it a challenging environment for the organisms and lowers the resilience of the ecosystem (Tomczak et al., 2013). Further, these special characteristics affect the suitability of biofouling management methods in the Baltic Sea (Korpinen et al., 2012). Fouling release coatings, for example, are not ice resistant and can therefore be used only in ice-free marine areas. Generally, the copper release of coating increases with increasing salinity (Valkirs et al., 2003), however, low salinity increases the toxicity of copper, since copper stays in its labile form. Therefore, the copper load in the low saline Baltic Sea can cause more severe consequences in the Baltic Sea than in other marine areas. Finally, the busy marine traffic to and from the Baltic Sea increases the NIS introduction risk and their spreading potential. The marine traffic from outside of the Baltic Sea causes a NIS introduction risk of new species while the internal marine traffic causes the risk of secondary spread (Ojaveer et al., 2017).

In this study, for modeling purposes, the Baltic Sea is divided into five sub-areas based on the salinity, the water temperature and the number of reported NIS in the area (Fig. 1). The division roughly follows the ICES (International Council for the Exploration of the Sea) subdivisions. The areas from west to east are (see Fig. 1 for sub-divisions indicated in the parenthesis): The southwestern Baltic (21-24), Baltic Proper (25-29, 28.1 excluded), The Gulf of Riga (28.1), The Gulf of Bothnia (30-31) and The Gulf of Finland (32). In addition, the eastern part of the North Sea is taken into account since the majority of the marine NIS introductions to the Baltic Sea originate from that area.

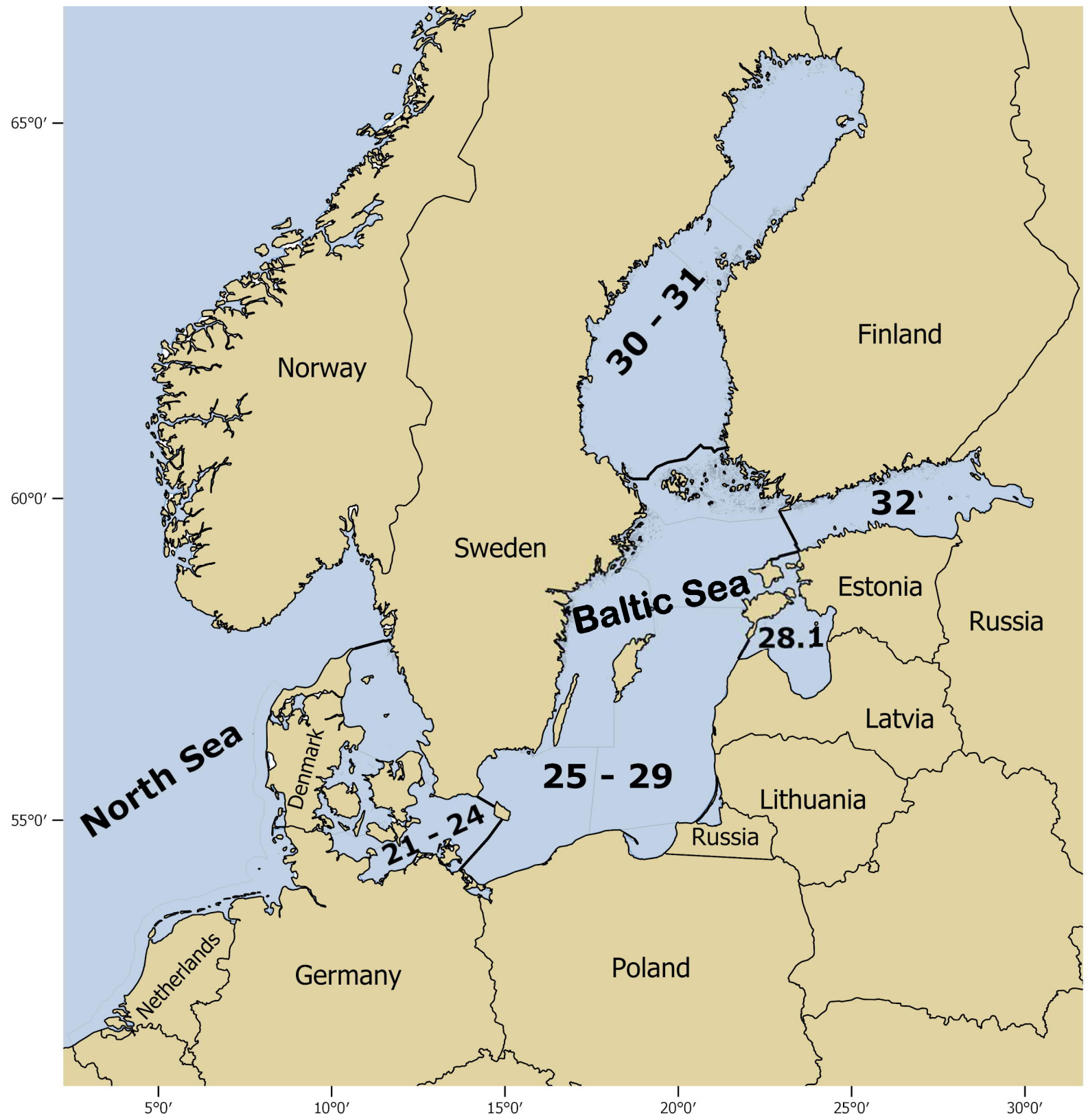

Figure 1. The study area division applied in the study, the North Sea and Baltic Sea that is further divided to five sub-area divisions following the boundaries and numbering of the ICES subdivisions. 21-24: The southwestern Baltic; 25-29: The Baltic Proper; 28.1.: The Gulf of Riga; 30-31: The Gulf of Bothnia; and 32: The Gulf of Finland.

## 2.2 Bayesian Network

We applied Bayesian Networks (BNs) to construct the decision support tool. BNs are probabilistic acyclic graphical models consisting of variables (nodes) and conditional dependencies (arrows) between them (Jensen, 2009). The strength of the dependencies is defined with the conditional probability tables (CPTs) (Bromley et al., 2005) where the probabilities are used to tell how certain things can occur (O'Hagan et al., 2006).

In BNs, CPTs can be defined using different information sources (qualitative or quantitative) such as expert or stakeholder knowledge (O'Hagan et al., 2006; Shaw et al., 2016; Salliou et al., 2017), literature (Lecklin et al., 2011) and observed data (Uusitalo et al., 2012) and combinations of these (Lehikoinen et al., 2013; Xue et al., 2017). This flexibility makes them highly useful in multi and interdisciplinary research. Further, BNs can cope with missing values and they are easy to update (Uusitalo, 2007). Therefore, BNs are useful in modeling environmental, complex problems where the data is often sparse and even missing.

An influence diagram (ID) is a generalization of a BN, allowing to model and solve decision-making problems under uncertainty (Nielsen and Jensen, 2009). An ID includes three types of nodes: random (uncertain) nodes, decision nodes (e.g. management strategies and options), and utility nodes to measure the utility (or loss) to be achieved by selecting the alternative decisions.

For management purposes, BNs can be applied either to infer the most probable effect given a cause or vice versa based on the gained evidence for the effect. For decisive purposes, the utility of a certain decision can be examined for instance in monetary terms (e.g. Helle et al., 2015). BN offers other benefits for analyzers too: The method allows complex problems to be represented graphically making the system easier to understand (Klemola et al., 2009). This is a particular advantage when cooperation is done with non-scientific stakeholders: in such a situation a visualized method containing the main features that are relatively easy to explain and comprise is especially important to ensure goal-oriented cooperation. Therefore, we choose to apply BN methods in this study where various data from different sources is used and a complex environmental problem is modelled to help non-scientist stakeholders to examine the biofouling management issue.

### 2.3 Data

As a starting point to construct the DST we utilized a conceptual influence diagram (CID) by Luoma et al. (2021), developed for structuring and visualizing the biofouling management problem in the Baltic Sea. We applied the CID to recognize the relevant information needed to solve the management problem such as "what kind of a management strategy would be optimal given the characteristics of the ship, its use and different environmental and economic aspects". Based on the availability and properties of the data, the CID structure was modified to create the DST capable of computational decision analysis.

To construct the decision support tool, we collected data from different sources: a database, literature, experts, stakeholders and through on-board measurements (Table 1). Face-to-face and phone interviews were made for the stakeholders representing shipping companies and an IWC company; a literature review was made considering the biofouling management; several on-board measurements were performed in three ships (7 periods) sailing in the Baltic Sea (Appendix A) and voyage data was recorded in five vessels; two workshops and four project meetings were held with the COMPLETE project (https://www.balticcomplete.com/) experts consisting of scientists, authorities and other specialists working on the topic. For further information considering the data collection see Luoma et al. (2021). A separate model was developed to estimate the case-specific NIS introduction risk (see Appendix B).

Table 1. Data collection method, timing, the quantity and use.

| Data collection method | Time | Quantity | Data used for structuring (S) or for quantifying (Q) the model |
|---|---|---|---|
| AquaNIS database | 2018-2021 | See Appendix B | Q |
| Face-to-face interviews | May-September 2018; March-October 2019 | 9 | S |
| Phone interviews | Autumn 2019 | 3 | S and Q |
| Literature review | 2018-2019 | 1 | S and Q |
| A model to estimate the probability of NIS introduction | 2020 | See Appendix B | Q |
| On-board measurements | 2018-2019 | 7 | Q |
| Voyage data recordings | 2018-2019 | See Appendix A | Q |
| Workshops | October 2018 (Helsinki, Finland); March 2020 (online) | 2 (15-20 participants in each) | S and Q |
| Project meetings | April 2018 (Riga, Latvia); December 2018 (Gothenburg, Sweden); April 2019 (Klaipeda, Lithuania); December 2019 (Jurmala, Latvia) | 4 (20-30 participants in each) | S |

# 3. Results

The decision support tool (DST) (Fig. 2) was constructed by using Hugin Researcher software (version 8.8; Madsen et al., 2005). The structure of the tool was discussed together with the project experts. The selection of the variables in the model was primarily based on their relevance in case-specific sustainable biofouling management and secondary on the possibilities to gain the data for the probability tables.

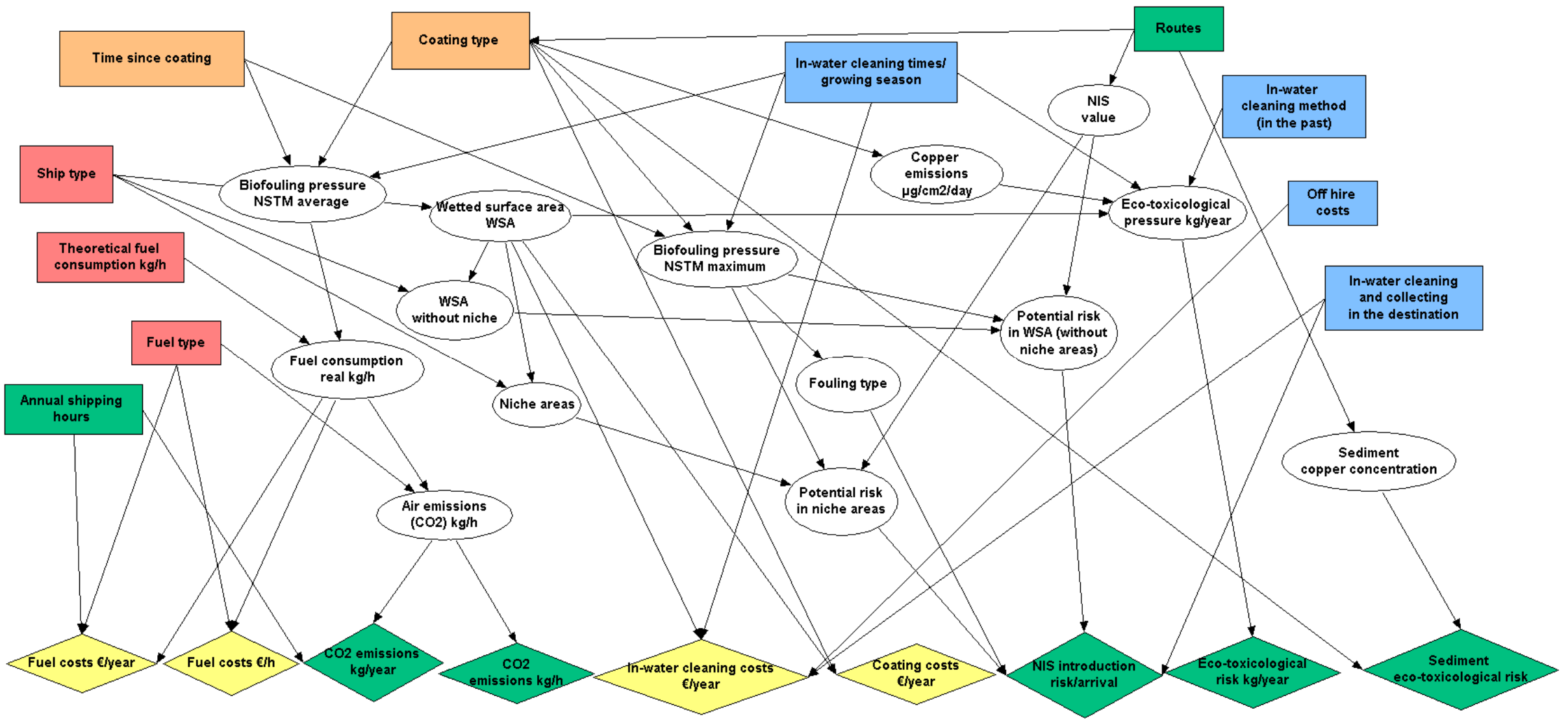

Figure 2. The DST with decision nodes (red rectangles = ship related; green rectangles = operational profile related; orange rectangles = coating related; blue rectangles = in-water cleaning related), random nodes (white ovals), utility nodes (yellow diamonds = costs; green diamonds = environmental impacts) and conditional dependencies (arrows) between the nodes.

The tool consists of 11 decision variables (rectangles), 9 utility variables (diamonds), 14 random variables (ovals) and their conditional dependencies (arrows). All the variables are described below but for further details see the Appendix C.

The decision variables enable the case specific biofouling management analysis. The end-user can select one option for each decision variable and hence test the effects of different combinations of management actions. The random variables represent the most relevant elements of the system through which the jointly made decisions impact on the utility variables. Each random variable has several possible states, the probabilities of which depend on the states of the preceding random and decision variables. The utility variables play a key role when the aim is to compare different decisions, as they present the consequences to be considered in the decision making process of the biofouling management strategy. The utility variables present the actual gains and losses of each scenario. Although the units of the utility variables are not commensurable, utilities and losses in different scenarios can be compared. All the utilities, except the NIS introduction risk and sediment ecotoxicological risk, are calculated annually. The NIS introduction risk presents the risk of each arrival to the port and thus depends on the annual arrivals. The sediment ecotoxicological risk depends on the current sediment copper concentration, copper released to the sediment from the coating and the coating type used: biocidal coating (BC) with IWC increases the risk while non-biocidal hard coating (HC) or fouling release coating (FR) does not. If the copper concentration in the area is high, it is important to use non-biocidal coatings and especially avoid IWC on biocidal coatings.

Decision variables

*Ship type* has six states: bulker, container, general cargo, passenger, RoRo and tanker. According to HELCOM report (2018), based on AIS data from the years 2006-2016, the most common large vessel-types in the Baltic Sea are cargo, tanker, passenger and container ships which account even 80 % of the traffic of the larger ships in the Baltic Sea registered by IMO (International Maritime Organization).

*Theoretical fuel consumption kg/ h* has interval states ranging from 1000 to 5000 kg/h. The user can set the suitable state based on the ship's unique characteristics.

*Fuel type* has states: light and heavy fuel oil. High quality light fuels burn cleanly and hence, produce less harmful emissions than heavy fuels. In addition, the $CO_2$ emissions per burnt ton of fuel depend on the type of fuel and its degree of refining (IMO, 2014). Fuel tariff levels favor heavy oils but due to the environmental regulations, the increasing proportion of ship-owners utilizes light fuels.

*Annual shipping hours* are related to the operational profile of the ship and are often established. Here intervals range from 1000 to 8760 h/year.

*Routes* have ten states: 10 different routes to, from and inside the Baltic Sea were considered. The routes were chosen based on their activity but also their dissimilarity was considered. Routes impact the NIS introduction risk and sediment eco-toxicological risk. In the Baltic Sea, the number of NIS varies between areas. Further, the environmental conditions between the departure and destination port (which together specify the route) may differ considerably, for which reason only certain NIS are able to spread and cause a NIS introduction risk. In addition, if the end-user has information on the areas of the highest NIS introduction risk and sediment eco-toxicological risk, it may affect his or her decision on where to perform the IWC.

*Time since coating* has states from 0-4 years. Usually the dry-docking, where old coating is replaced, takes place every 3-5 years. The condition of the hull and its coating affect the biofouling, since the more worn the coating the higher is the biofouling level. In addition, the IWC methods applied affect the condition of the hull. For simplicity, here the *Time since coating* determines the wear of the coating.

*Coating type:* The most common coating types used in the Baltic Sea were chosen: non-biocidal hard coating (HC); biocidal self-polishing coating (BC); and fouling release coating (FR). HC must be cleaned regularly due to the fast biofouling but BC containing copper can be cleaned much more seldom, whereas FR coating can be completely without IWC in the Baltic Sea. However, FR coating does not resist ice friction and thus can only be used on ice-free routes.

*In-water cleaning times/growing season:* The IWC is performed in the Baltic Sea only during the growing season; from April to the end of September. The states are 0, 2, 6 and 12 meaning not performed at all, twice a year or once-twice a month during the growing season.

*In-water cleaning method (in the past)* has two states: soft brushes and hard brushes. If the in-water cleaning is performed, it is usually conducted with nylon or steel brushes. When the hull biofouling is soft, nylon brushes causing less damage to the coating can be used. However, when hard fouling occurs, steel brushes causing damage to the coating and releasing more copper must be used. Sometimes IWC is not performed e.g. when FR is used (Stakeholder feedback 2019). Here only the increased copper releasing was considered, not the damages caused to the coating.

*Off hire costs* has three states: none, 1 day, 2 days. Here the off hire cost is assumed to be 20 000 €/day. While passenger ships can be cleaned in the ports during loading, for safety reasons tankers are not allowed to perform in-water cleaning where loading occurs and thus cannot avoid the off hire costs. Sometimes other ship types have off hire costs as well.

*In-water cleaning and collecting the material in the destination* has three states: IWC with collecting, IWC without collecting and no IWC. The NIS introduction risk of the states was estimated with an expert. According to the expert, there are many different species (in soft and hard fouling) with different tolerance and different viability and thus it is hard to put these in general order. The order usually depends on the species attached to the hull. However, generally the best option is IWC with the capture device (filters with the smallest holes and the water with organic material collected and filtered). This way the number of fouling species to be introduced to the area is the lowest. The worst option is in-water cleaning without any capture. This option removes the species from the hull to the water resulting in surviving and viable organisms. Thus the second best option is “no in-water cleaning”. This means that the species are not removed from the hull but still some of them can be detached by themselves and at least some species can release offspring to the area. There are knowledge gaps considering these options for hard and soft fouling, however, the order can be assumed to be the same for the soft and hard fouling. In addition, the data in Morrisey et al. (2013) was used to formulate the probability distributions in more detail.

**Random variables**

*Biofouling pressure NSTM average* and *Biofouling pressure NSTM maximum* illustrate the biofouling level of the hull affecting the fuel consumption and NIS introduction risk. Biofouling level is often determined by a visual inspection and categorized according to Naval Ships' Technical Manual (NSTM) rating of the US Navy (US Navy, 2006), ranging from 0 to 100 (soft fouling: light–heavy slime 10–30, hard fouling: small–heavy calcareous fouling and weed 40–90). The biofouling pressure NSTM average will be used for the fuel consumption calculations and the biofouling pressure NSTM maximum for the NIS introduction risk calculations representing the worst-case scenario of the biofouling level at a certain time. The probability distributions for biocidal coating were calculated based on data published in Pagoropoulos et al. (2017). In addition, since one of the authors performed the fuel consumption measurements in passenger ships sailing in the Baltic Sea (Appendix A), he elicited based on the on-board measurements the probability distributions for hard coating and modified the biocidal coating distribution to better fit the Baltic Sea environment. Based on the interviews performed, fouling release coating does not need IWC in the Baltic Sea and therefore their biofouling pressure was estimated very low.

*Wetted surface area (WSA), WSA without niche areas* and *Niche areas* represent the area of the hull being fouled during shipping. Wetted surface area (WSA) is a standardized measure of the maximum submerged surface of a hull that can potentially be fouled (Moser et al., 2017). Here the probability distribution for the WSA with different ship types was formulated by using the data of all the ships coming to Polish seaports in 2018.

Niche areas are often more heavily fouled than other WSA and therefore they are considered separately. The share of different niche areas varies between ship types. The data was received from Moser et al. (2017).

*Fuel consumption real kg/h* is based on the theoretical fuel consumption and the biofouling level increasing the fuel consumption. The effect of a certain biofouling pressure on fuel consumption was estimated based on the ship fuel consumption measurements.

*Air emissions* ($CO_2$) *kg/h* depends on the fuel consumption and fuel type being approximately a triple compared with the fuel consumption.

*Fouling type* is dependent on the biofouling level: here it is assumed that with a high biofouling level the fouling type is hard (40-100 NSTM) and with a low level (0-30 NSTM) the type is soft.

*NIS value* represents the possible number of NIS able to be introduced from the departure port to the destination port. The NIS number is calculated with a model code based on the collected data of the NIS in a certain area and their ability to attach onto the ship's hull and survive the trip when the salinity changes (Appendix B).

*Potential risk in niche areas* and *Potential risk in WSA (without niche areas)* are variables impacting the NIS introduction risk. The potential risk consists of the ship's fouled area, the biofouling level and the number of NIS. Ten in-water diver inspections were used to compare the biofouling level between niche areas and the main hull (Dinis Reis Oliveira, personal communication). According to the results, the niche areas can be even 2-3 times more heavily fouled than the main hull. Here we used the multiplier 2.6.

*Copper emissions µg/cm2/day* increase the risk of ecotoxic response in aquatic organisms. The copper is released if the biocidal coatings are used. We used an expert to estimate the copper emissions. According to the expert, the copper emission from the ship's hull treated with biocidal coating in the Baltic Sea is on average 7 µg/ cm2/ day. This estimation was used for the average copper emission rate (i.e. the rate when the ship is sailing in the Baltic Sea).

*Eco-toxicological pressure kg/year* refers to the amount of copper released to the Baltic Sea from a ship with a certain biofouling management method. Generally, a higher emission peak is followed after IWC. According to the expert, estimations for the copper emission rate after IWC in the Baltic Sea are not available, however, the rates found in the literature, such as Morrisey et al. (2013) are applicable to the Baltic Sea. Therefore, following estimations were used: with soft methods, the higher peak (seven days) is 12 µg/ cm2/ day and with hard methods the higher peak is (seven days) 25 µg/ cm2/ day.

*Sediment copper concentration* was based on the sediment copper data available to all study areas in the Baltic Sea except the Gulf of Riga (Gulf of Finland: Gubelit et al., 2016, Vallius, 2012; South-western Baltic: ICES, 2014, Nikulina, 2008; Other study areas: ICES 2014). For comparability, surface sediment samples sieved to < 2 mm were used. Copper concentrations were interpolated to all study areas within the Baltic Sea using inverse distance weighting and zonal statistics tools in QGIS (v.3.4.11). The sediment copper threshold value was defined according to environmental quality standards for copper in marine sediment calculated by Sahlin and Ågerstrand (2018). Accordingly,*t* applying a threshold value of 52 mg/kg that considers 5 % organic carbon content, the copper content was classified into low and high.

Utility variables

*Fuel costs* can be considered on an annual (€/year) or hourly (€/h) basis and are one of the main costs of the shipping. Light fuel oil is more expensive than heavy fuel oil.

*$CO_2$emissions* can also be considered on an annual (*kg/year)* or hourly (kg/h) basis and are connected to the *fuel type* and *fuel consumption.* The light fuel is cleaner producing less toxic emissions than heavy fuel.

*In-water cleaning costs* €/year depend on the IWC times/growing season and the WSA. Usually only the sides of the hull are cleaned as the biofouling diminishes rapidly when the light conditions dim. Here we assume that 40 % of the total WSA is cleaned and the cost is 3€/$m^2$ which is based on the cost estimations received from the interviewed IWC company. The IWC costs with devices collecting the material are estimated to be 50 % higher than without the devices.

*Coating costs* €/year consist of immersed hull surface treatment costs and durability of the selected coating. In addition to the use of surface treatment agents and the amount of work required, all preliminary work needed especially when changing the hull treatment method increases final costs cumulatively. The durability and renewing of the treatment produces not only periodic direct costs but also indirect costs when the vessel is out of service. Here we assume that the coating is renewed every fifth year. It was hard to get data considering the different coating costs so we assumed the application costs are 2€/$m^2$ for all coating types. The estimated coating costs are presented below in Table 2.

Table 2. The estimated coating costs of different coating types.

| Coating | €/m$^2$ |
|---|---|
| Self-polishing biocidal coating (BC) | 20 |
| Hard coating (HC) | 30 |
| Fouling release coating (FR) | 50 |

*NIS introduction risk/arrival* depends on the potential attachment risk onto the hull and on the fouling type. Some NIS may have less severe impact or do not have verifiable effects; however, in this study, we follow the viewpoint of HELCOM (2007), which states that all NIS introductions in the Baltic Sea should be avoided. Additionally, the biofouling pressure of the WSA, the fouling type (soft or hard) and the possible IWC method used in the port influence the risk. In addition, niche areas such as bilge wells can carry larger amounts of biofouled biota and can be difficult to clean periodically. IWC releases the biofouling from the hull but with effective capturing devices the NIS introduction risk can be minimized. It should be acknowledged that even without IWC, species can detach from the hull or spread gametes causing a NIS introduction risk.

*Ecotoxicological risk* kg/year refers to the copper emissions released from the coating. The ecotoxicological risk is calculated as kg/year and is dependent on the coating type, WSA, IWC /growing season and method. Although the routes with varying environmental conditions can represent different ecotoxicological risk characteristics, here the whole Baltic Sea is considered as a uniform area.

*Sediment eco-toxicological risk* is estimated based on the current sediment copper concentration in a certain area and the copper released to the water from the biocidal coating. Especially marinas and harbours may contain relatively high, ecologically harmful copper concentrations (Eklund et al., 2010).

## 4. Discussion

We constructed a quantitative Bayesian network decision support tool (DST) allowing to study multidimensionally and case-specifically the complex biofouling management problem in the Baltic Sea. The DST visualizes the biofouling management phenomena and the multiple perspectives, increases the systemic understanding and awareness about different aspects and their interlinkages, and provides an interface to understand and study the subject. The tool is useful for anyone interested in this topic (See the user manual in Appendix D and the model code in Appendix E) but especially for decision makers and authorities working with biofouling management issues. A selection method for the best coating for a certain ship has been missing (Uzun et al., 2019) but this tool offers support in choosing the optimal management options and comparing different biofouling management strategies. In addition, the tool allows the user to study two main biofouling management methods (coating and IWC) together providing the risks and costs related to monetary costs (biofouling and fuel consumption), $CO_2$ emissions, ecotoxicological risk (copper load) and NIS introduction risk. Thus, it is the first tool of this kind considering widely multiple perspectives and allowing case-specific analyses in a visual format.

According to the interviews (Luoma et al. 2021), only recently the shipping industry has started to be more aware of the slightest biofouling layer and its impact increasing the annual fuel costs. The slowly growing level of fuel consumption is often overlooked (Pagoropoulos et al., 2017). The regular cleaning of a slightly biofouled hull may have been seen as unnecessary if attention has not been paid on the annual fuel consumption and $CO_2$ emission increase, the increased NIS introduction risk and the required harder cleaning process, due to the more biofouled hull, causing coating damages. The novel DST supports the optimal biofouling management by helping to realize the annual increase in fuel consumption even with a low biofouling level and revealing the ecological risks connected to that.

In the future, the tool can be developed with a user-friendly interface and platform-free software solutions to ease the use of the tool especially among non-research stakeholders. The biofouling management costs and fuel costs change temporarily and spatially. Therefore, the tool can be developed to make it easier to modify the costs by the user himself/herself.

The visual format of the DST is advantageous since visuals can produce a useful presentation in the world with a huge information load (See van der Hoorn, 2020 and references therein), increase open communication between different groups (See Carriger et al., 2018) and even develop the systemic understanding about the issue. A visual tool like this can support transparent decision making and communication by making visible the realized connections between different variables and all the divergent aspects considered in the tool.

Commonly, the ship biofouling management decisions are made from an economic perspective (Schultz, 2004; Schultz et al., 2011), especially fuel costs are emphasized, and less attention is paid on the ecological or social perspective. This tool widens the perspective and shows that ecological risks such as ecotoxicological risk and sediment ecotoxicological risk are low when non-biocidal coatings are used. The tool shows that the optimal biofouling management often consist of using a biocidal-free coating with regular in-water cleaning and filtering devices collecting the released material. With this kind of strategy both the eco-toxicological risk and the NIS introduction risk is low. In addition, savings in the fuel costs due to the lower biofouling level makes it economic as well. On the other hand, the best biocidal-free coating type is case-specific. In the southern Baltic Sea with ice-free conditions, FR is an optimal choice but in the areas with partial ice-cover in winter FR cannot be used and then HC is often the best option. The exceptions are tankers, which need to perform IWC, due to the safety reasons, in other ports than where the loading occurs. Therefore, tankers have less frequent possibilities for IWC and higher IWC costs than other ship types. Thus, the BC can be the optimal method for tankers in some cases. However, even with the BC the biocide content should be adjusted to the environment and be only as high as needed. Therefore, there is still a need for better IWC methods and effective but environment-friendly coatings to reach better biofouling management in the Baltic Sea.

## Acknowledgements

This research was prepared within the project COMPLETE - Completing management options in the Baltic Sea region to reduce risk of invasive species introduction by shipping [#R069]. The COMPLETE project is co-financed by the European Union's funding Programme Interreg Baltic Sea Region (European Regional Development Fund). The authors would like to thank all the interviewees for their participation, all the COMPLETE project partners for their comments and support and especially Dr. Dinis Reis Oliveira and Lauri Nevalainen for their valuable input and expertise. The work of I. Helle was funded by the Helsinki Institute of Sustainability Science (HELSUS), University of Helsinki. The work of A. Lehikoinen was funded by the Strategic Research Council at the Academy of Finland (project WISE; decision no. 312625).

# Supporting Information

On-board measurements and voyage data recording has been performed in five ships with the aim of finding out the measurable effect of immersed hull biofouling of regularly cleaned ships operating in the Baltic Sea. In table A1, the voyage data recording periods have been illustrated. All the periods have been chosen to focus on the summer season because hull contamination is higher in the late summer when the sea water is warm and a lot of light is available also in the northern latitudes. All the ships have been operated by associated partners of the COMPLETE project. Ships A, B and C are sister ships, ships D and E can be classified as similar but still containing different structures. Because the environmental conditions always vary, recordings have been done over long periods. Thus, recorded data includes several periods whereby the conditions before and after the immersed hull cleanings are approximately comparable.

Table A1. The voyage data recording periods

| Period | Ship | Time |
|---|---|---|
| 28 - 31 May 2018 | ROPAX (Ship A) | 3 days |
| 20 Jul - 3 Sep 2018 | ROPAX (Ship A) | 37 days |
| 1 - 8 Aug 2018 | ROPAX (Ship B) | 7 days |
| 14 - 21 Aug 2018 | ROPAX (Ship B) | 7 days |
| 30 Jun - 14 Jul 2019 | ROPAX (Ship B) | 14 days |
| 22 Apr - 12 Oct 2019 | ROPAX (Ship C) | 173 days |
| 1 Mar - 13 Oct 2019 | ROPAX (Ship D) | 226 days |
| 24 - 26 Apr 2019 | ROPAX (Ship E) | 2 days |
| 20 - 22 Aug 2019 | ROPAX (Ship E) | 2 days |

Recorded voyage data variables and sources are shown in table A2. The speed of the ship is an important variable to determine the ship's resistance but unfortunately the technology allows only recording speed over ground. Course and position are needed when the effect of prevailing weather conditions is noticed. Fuel consumption, pitches and shaft rotation speed indicates operating conditions of the engines and data given by shaft torsion meters produces the best available data of power for the propellers. However, the latter was only available from the L3 Valmarine system. Trim and draught varies depending on the loading conditions and trim conditions also have an effect on drag, depending on the type of vessel. The voyage-specific DWT is obtained from the NAPA hub. The regional wind condition data have been offered from coastal meteorological stations and current estimations are given by the BSH.

Table A2. The voyage data variables

| Variable | Source |
|---|---|
| SOG | L3 Valmarine APIS / BlueFlow / AIS |
| COG | AIS |
| Position | AIS |
| Fuel consumption PP | L3 Valmarine APIS / BlueFlow |
| Fuel consumption STBD | L3 Valmarine APIS / BlueFlow |
| Trim | L3 Valmarine APIS / BlueFlow |
| Draught | L3 Valmarine APIS |
| Pitch PP | L3 Valmarine APIS / BlueFlow |
| Pitch STBD | L3 Valmarine APIS / BlueFlow |
| Rotation speed PP | L3 Valmarine APIS / BlueFlow |
| Rotation speed STBD | L3 Valmarine APIS / BlueFlow |
| Shaft power PP | L3 Valmarine APIS |
| Shaft power STBD | L3 Valmarine APIS |
| DWT | NAPA |
| Wind speed | Coastal weather stations |
| Wind direction | Coastal weather stations |
| Current estimations | BSH |

The aim of the on-board emission measurement sessions was to determine the emission levels of ships under investigation. At the beginning of the project, the Thetis-Mrv database launched by EMSA was still under construction. However, this database focuses on annual emission levels thus in-service emission levels for this purpose on-board measurements provide more accurate information. The emission measurements have been carried out on the same ships but for shorter periods shown in the table A3.

Table A3. The emission measurement session periods

| Period | Ship | Time |
|---|---|---|
| 28 - 31 May 2018 | ROPAX (Ship A) | 3 days |
| 21 - 24 Aug 2018 | ROPAX (Ship A) | 3 days |
| 24 - 26 Apr 2019 | ROPAX (Ship E) | 2 days |
| 22 - 24 May 2019 | ROPAX (Ship D) | 2 days |
| 20 - 22 Aug 2019 | ROPAX (Ship E) | 2 days |
| 2 - 5 Sep 2019 | ROPAX (Ship A) | 3 days |
| 23 - 25 Sep 2019 | ROPAX (Ship D) | 2 days |

Measured emission gaseous components are shown in table A4. During measurements, levels of sulfur dioxide, nitrogen oxides, carbon monoxide and dioxide, residual oxygen, exhaust gas flow and temperature have been observed. Most of the components are measured with a Horiba PG350 equipment following standards set out in column but flow rate and temperature needed separate equipment.

Table A4. The emission variables

| Variable | System |
|---|---|
| SO2 | HORIBA PG350 (CEN/TS 17021:2017) |
| NOx | HORIBA PG350 (SFS-EN 14792) |
| CO | HORIBA PG350 (SFS-EN 15058) |
| CO2 | HORIBA PG350 (ISO 12039) |
| O2 | HORIBA PG350 (SFS-EN 14789) |
| Flow | Mikor TT570SV |
| Temperature | Fluke thermo |

This approach leads to a typical mass data problem including all five describing characteristics: volume, variety, velocity, value and veracity. The voyage data is collected from several sources, where time step, format and accuracy varies, and data sets give primary or indirect information based on the source. Mainly, the data collection is an automatic process and recorded data is not intended to be utilized for analyzing research problem of this kind. This means that always when analyzing mass data, the actual framework must be considered to avoid erroneous conclusions.

The data analysis defining extra drag caused by the hull biofouling was performed as follows: Biofouling grows fastest in the late summer, so the data sets from mid-July to early September were selected for analysis. The route and the ship have been selected so that the ship is sailing in high sea areas where the depth is enough in relation to her size and actual running speed for minimizing steering and shallow water effects. In addition, on the high seas, currents tend to be more linear than on the archipelago coast. However, in order to be able to rely on the weather data provided by the coastal weather stations, the route should run relatively close to the coast. The loading conditions should be close to each other on comparable voyages as well. The actual hull cleaning days set a threshold for a clean and biofouled hull. The biofouling level is approximated based on reports given by scuba divers.

The prevailing weather in the area is observed on the basis of the data provided by coastal weather stations. Only the etaps where weather conditions are calm are selected for the analysis. Additional certainty is obtained by calculating theoretical efficiency for the ship because under comparable conditions the efficiency should be at the same level. Once the comparable etaps have been selected before and after the hull cleanings, the data classification has been performed by using naïve bayes algorithm. It is a quite simple and efficient model including strong independence assumption between all the input variables. Despite this assumption, the classifier still works quite reliably.

In particular, the analyzes based on the voyage data at least in the theoretical framework could also be feasible by shipowner. One of the aims was to utilize tools and methods where this goal could be achieved. The ship hulls included in this study have been treated by hard coat type epoxy coatings, which are suitable for operating in icy conditions but get biofouled during summer. With this kind of coatings, the ship drag increasing at the first month after cleaning seems to be close to 2-4 %. However, the growth of biofouling does not occur linearly and often during the second month biofouling adheres to the hull much more tightly. Later, the number of hard-shelled organisms also increases. Therefore, the extra drag and increased fuel consumption is only one aspect of technical disadvantages caused by the biofouling: Tight-fitted biofouling is challenging to clean off and the process often requires utilization of steel brushes that also damage coating surfaces. Regular cleanings also give savings from this point of view.

The model description to estimate the probability of NIS introduction in each route

We constructed a BN model to estimate the NIS value, i.e. the potential number of NIS on ships' hull that can be spread from the departure area to the arrival area. As the decision support tool follows the viewpoint of The HELCOM Baltic Sea Action Plan (HELCOM, 2007), i.e. the precautionary principle, all the species that have been found at the departure area was taken into the account, even if they have been already found in the arrival area. Even if some of these species have already been found in the arrival area, it may still create enhanced risk on the area, as the species might have not been found in the specific area where the ship arrives and stays for a longer time.

In this model calculating the NIS value, salinity is considered as a sole critical environmental factor limiting the NIS introduction to the "new" area. Temperature and salinity have been found to be significant environmental factors that may limit the spread of the species in the Baltic Sea (Ojaveer et al., 2010). However, especially in brackish seas salinity is the most significant range limiting factor regulating species distribution and colonization (Cognetti and Maltagliati, 2000; Paavola et al., 2005; Villnäs, & Norkko, 2011; Virta et al., 2020). Specifically in the northern latitudes, climate change is predicted to result in more precipitation and subsequent fresh-water runoff in the Baltic Sea (HELCOM, 2007; Leppäranta and Myrberg, 2009), and hence reduced salinity can be more significant environmental factor than temperature in regulating the introduction of NIS to the new areas (Somero, 2012; Holopainen et al., 2016).

In the Baltic Sea, the salinity of the surface water layers varies by region. The hydrographical characteristics, surface (0 m) salinity in the study areas between years 2010 – 2018 was analyzed using Copernicus database (www.copernicus.eu) rasters. Spatial analysis was carried out using QGIS (v 3.4.11) spatial analysis tools (multiband zonal statistics) to calculate the monthly mean, minimum and maximum of salinity. For this model, only mean values were used.

The model quantifies the uncertainty around the minimum and maximum salinities in each area. However, some limitations should be noted when interpreting the results of the model calculating the NIS value. The model overlooks the uncertainty of the NIS data as it does not consider how the NIS have been monitored (how frequently or which methods have been used, how widespread the species is) in each country. Salinity tolerances of the species are taken as fixed numbers. The length of the route can also influence how many species can survive to the arrival area as the longer the route is, the higher is the risk that the species or part of it can fall. This may pose a higher risk along the way but correspondingly reduce the risk at the arrival area.

The list of NIS used in the analysis, their introduction status in the study areas and salinity preferences are presented in Table B1.

Table B1. The list of non-indigenous species used in the analysis, their salinity preferences and introduction status in the study areas (whether the species is present or not). Sal_min_tol = salinity minimum tolerance; Sal_max_tol = salinity maximum tolerance; GoB = The Gulf of Bothnia; The Gulf of Finland; GoR = The Gulf of Riga; BP = Baltic Proper; SWB = The southwestern Baltic; NS = The eastern part of the North Sea.

| Species | Sal_min_tol | Sal_max_tol | Present_GoB | Present_GoF | Present_GoR | Present_BP | Present_SWB | Present_NS |
|---|---|---|---|---|---|---|---|---|
| Acrochaetium catenulatum | 15 | 26 | No | No | No | No | No | Yes |
| Aglaothamnion halliae | 17 | 30 | No | No | No | No | No | Yes |
| Alitta succinea | 0.14 | 80.0 | No | No | No | No | Yes | Yes |
| Amathia gracilis | 15 | 30 | No | No | No | No | Yes | Yes |
| Antithamnionella ternifolia | 19.0 | 35.0 | No | No | No | No | No | Yes |
| Antithamnionella spirographidis | 19 | 35 | No | No | No | No | No | Yes |
| Aplidium glabrum | 19.0 | 35.0 | No | No | No | No | No | Yes |
| Arcuatula senhousia | 18.0 | 35.5 | No | No | No | No | No | Yes |
| Ascidiella aspersa | 18.0 | 40.0 | No | No | No | No | Yes | Yes |
| Austrominius modestus | 14.0 | 42.8 | No | No | No | No | No | Yes |
| Balanus amphitrite | 10.0 | 52.0 | No | No | No | No | No | Yes |
| Bonnemaisonia hamifera | 10 | 38 | No | No | No | No | Yes | Yes |
| Botrylloides violaceus | 15.0 | 34.0 | No | No | No | No | No | Yes |
| Botryllus schlosseri | 14.0 | 44.0 | No | No | No | No | Yes | Yes |
| Bugula neritina | 14.0 | 40.0 | No | No | No | No | No | Yes |
| Bugulina simplex | 18 | 30 | No | No | No | No | No | Yes |
| Bugulina stolonifera | 15 | 40 | No | No | No | No | No | Yes |
| Caprella mutica | 11.0 | 40.0 | No | No | No | Yes | Yes | Yes |
| Carcinus maenas | 1.4 | 52.0 | No | No | No | Yes | Yes | Yes |
| Chelicorophium curvispinum | 0.5 | 8.8 | No | Yes | Yes | Yes | Yes | Yes |
| Chelicorophium robustum | 0.5 | 8.8 | No | No | No | Yes | No | Yes |
| Codium fragile subsp. fragile | 12.0 | 48.0 | No | No | No | No | Yes | Yes |
| Codium fragile | 17.5 | 40 | No | No | No | No | No | Yes |
| Codium fragile scandinavicum | 17.5 | 40 | No | No | No | No | No | Yes |
| Colpomenia peregrina | 15 | 30 | No | No | No | No | No | Yes |
| Corbicula fluminalis | 0 | 50 | No | No | No | No | No | Yes |
| Corbicula fluminea | 0.0 | 24.0 | No | No | No | No | No | Yes |
| Corella eumyota | 19.0 | 35.0 | No | No | No | No | No | Yes |
| Corynophlaea verruculiformis | 28 | 31 | No | No | No | No | No | Yes |
| Crassostrea gigas | 3.0 | 56.0 | No | No | No | No | Yes | Yes |
| Crassostrea virginica | 0.0 | 40.0 | No | No | No | No | Yes | Yes |
| Crepidula fornicata | 5.0 | 40.0 | No | No | No | No | Yes | Yes |
| Dasya baillouviana | 10.0 | 30.0 | No | No | No | No | Yes | Yes |
| Dasysiphonia japonica | 10.0 | 40.0 | No | No | No | No | Yes | Yes |
| Diadumene cincta | 18 | 40 | No | No | No | No | No | Yes |
| Diadumene lineata | 7 | 70 | No | No | No | No | No | Yes |
| Didemnum vexillum | 20.0 | 45.0 | No | No | No | No | No | Yes |
| Dikerogammarus haemobaphes | 0 | 0.5 | No | No | No | Yes | Yes | Yes |
| Dikerogammarus villosus | 0 | 24 | No | No | Yes | Yes | Yes | Yes |
| Diplosoma listerianum | 25.0 | 34.0 | No | No | No | No | No | Yes |
| Dreissena rostriformis bugensis | 0.0 | 5.0 | No | Yes | No | Yes | No | Yes |
| Ensis leei | 15.0 | 35.0 | No | No | No | No | Yes | Yes |
| Ficopomatus enigmaticus | 0.2 | 45.0 | No | No | No | No | Yes | Yes |
| Fucus distichus subsp. evanescens | 10 | 40 | No | No | No | No | Yes | Yes |
| Garveia franciscana | 1.0 | 35.0 | No | No | No | No | No | Yes |
| Gonionemus vertens | 14.0 | 34.0 | No | No | No | No | Yes | Yes |
| Gracilaria vermiculophylla | 8.5 | 60.0 | No | No | No | No | Yes | Yes |
| Grateloupia turuturu | 12.0 | 52.0 | No | No | No | No | No | Yes |
| Hemigrapsus sanguineus | 0.0 | 48.0 | No | No | No | No | Yes | Yes |
| Hemigrapsus takanoi | 0.0 | 40.0 | No | No | No | No | Yes | Yes |
| Hydroides dianthus | 28.0 | 50.0 | No | No | No | No | No | Yes |
| Hydroides elegans | 15.0 | 42.0 | No | No | No | No | No | Yes |
| Ianiropsis serricaudis | 17 | 35 | No | No | No | No | No | Yes |
| Incisocalliope aestuarius | 10 | 33 | No | No | No | No | No | Yes |
| Hemigrapsus penicilliatus | 7 | 35 | No | No | No | No | No | Yes |
| Jassa marmorata | 12.0 | 40.0 | No | No | No | No | Yes | Yes |
| Laonome xeprovala | 0.0 | 7.0 | Yes | Yes | Yes | Yes | No | No |
| Limnomysis benedeni | 0.0 | 14.0 | No | No | No | Yes | No | Yes |
| Limnoria lignorum | 6.0 | 50.0 | No | No | No | No | No | Yes |
| Limnoria quadripunctata | 18.0 | 48.0 | No | No | No | No | No | Yes |
| Melanothamnus harveyi | 5.0 | 34.0 | No | No | No | No | No | Yes |
| Melita ntida | 0 | 35 | No | No | No | Yes | No | Yes |
| Molgula manhattensis | 10.0 | 40.0 | No | No | No | No | Yes | Yes |
| Monocorophium sextonae | 25 | 42 | No | No | No | No | No | Yes |
| Myriophyllum heterophyllum | 0.0 | 33.0 | No | No | No | No | No | Yes |
| Mytilicola intenstinalis | 5 | 34 | No | No | No | No | No | Yes |
| Mytilopsis leucophaeata | 0.0 | 32.0 | Yes | Yes | No | Yes | Yes | Yes |
| Mytilus galloprovincialis | 10.0 | 38.0 | No | No | No | No | No | Yes |
| Nemopsis bachei | 0 | 23 | No | No | No | No | No | Yes |
| Obesogammarus crassus | 0.5 | 8.8 | No | No | No | Yes | Yes | Yes |
| Palaemon longirostris | 0.0 | 34.0 | No | No | No | Yes | No | No |
| Palaemon macrodactylus | 0.6 | 48.0 | No | No | No | Yes | Yes | Yes |
| Perophora japonica | 18 | 40 | No | No | No | No | No | Yes |
| Petricolaria pholadiformis | 7.5 | 35.0 | No | No | No | No | Yes | Yes |
| Polysiphonia brodiei | 5.0 | 35.0 | No | No | No | No | Yes | Yes |
| Pontogammarus robustoides | 0.0 | 23.0 | No | Yes | Yes | Yes | No | No |
| Rapana venosa | 12.0 | 32.0 | No | No | No | No | No | Yes |
| Rhithropanopeus harrisii | 0.4 | 40.0 | No | Yes | Yes | Yes | Yes | Yes |
| Sargassum muticum | 10.0 | 40.0 | No | No | No | No | Yes | Yes |
| Sinelobus vanhaareni | 4.9 | 7.3 | No | Yes | Yes | Yes | Yes | Yes |
| Smittoidea prolifica | 19.0 | 35.0 | No | No | No | No | No | Yes |
| Styela clava | 10.0 | 36.0 | No | No | No | No | Yes | Yes |
| Teredo navalis | 2.0 | 40.0 | No | No | No | No | Yes | Yes |
| Theora lubrica | 28 | 35 | No | No | No | No | No | Yes |
| Tharyx killariensis | 29 | 35 | No | No | No | No | No | Yes |
| Tricellaria inopinata | 20.00 | 35.0 | No | No | No | No | No | Yes |
| Ulva pertusa | 17.0 | 35.0 | No | No | No | No | No | Yes |
| Undaria pinnatifida | 6.0 | 34.0 | No | No | No | No | No | Yes |
| Victorella pavida | 0 | 36 | No | No | No | No | No | Yes |

The species introduced to a certain study area were determined using AquaNIS database (http://www.corpi.ku.lt/databases/index.php/aquanis). The search criteria was based on species introduced to study area ports and their capability to fouling (either hull or niche areas). Additional status of an introduction of a certain species on study area, if information from a port was not available, was checked manually from AquaNIS, literature as well as GBIF database (www.gbif.org). Later, the list of species was revised by the project NIS experts. Species that were either not factually foulers, introduced to all study areas or their population status was unknown, were excluded from the analysis.

NIS salinity preferences were determined using species' minimum and maximum tolerated values. Thus, the salinity interval represents the condition where a species is able to survive. The minimum and maximum salinity values were generally obtained from AquaNIS, however, complementary information was searched from other databases. In addition, for a certain species with no available value, either other physiologically valid information, e.g. maximum salinity for growth, or value from a species in the same family was used.

The model specification begins by assuming the minimum $x_{i,k}$ and maximum $y_{i,k}$ salinity by each data point *i* within each area *k* follow a normal distribution (N), (Eq. 1 and 2):

$$x_{i,k} \sim N(\mu_{x,k}, \sigma_{x,k} = C\mu_{x,k}) \quad (1)$$
$$y_{i,k} \sim N(\mu_{y,k}, \sigma_{y,k} = C\mu_{y,k}) \quad (2)$$

where $\mu_{x,k}$ and $\mu_{y,k}$ represent priors for the mean minimum and maximum salinities (the expected maximum salinities) of the area *k* and $\sigma_{xk}$ and $\sigma_{y,k}$ represent the standard deviation of the minimum and maximum salinities within each area *k*. The coefficient of variation C is assumed equal in all areas $C \sim LN(mean = 0.1, sd = 0.1)$. The prior distributions for the $\mu_{x,k}$ and $\mu_{y,k}$ follow a normal distribution (Eq. 3 and 4):

$$\mu_{x,k} \sim N(\nu_x, \sigma_x^2) \quad (3)$$
$$\mu_{y,k} \sim N(\nu_y, \sigma_y^2) \quad (4)$$

where $\nu_x$ and $\nu_y$ represent the means of these mean minimum and maximum salinities, and $\sigma_x^2$ and $\sigma_y^2$ represent the variance over minimum and maximum salinities at each area. Table B2 shows all the prior distributions used in the model.

To calculate the probability of species *s* introduction to the arrival area, we used the step-function that evaluates to 1 if the argument ≥ 0, and 0 otherwise. Our example code is provided below.

Table B2. Model parameters and their prior distributions. *Beta* = beta-distribution; *N* = normal distribution; *Unif* = uniform distribution.

| Abbreviations | Explanation | Prior distribution |
|---|---|---|
| C | The coefficient of variation | $C \sim LN(mean = 0.1, sd = 0.1)$ |
| $\boldsymbol{\sigma_x^2}$ | Variance over minimum salinities at each area | $\sigma_x^2 \sim Unif(0,10)$ |
| $\boldsymbol{\sigma_y^2}$ | Variance over maximum salinities at each area | $\sigma_y^2 \sim Unif(0,10)$ |
| $\boldsymbol{\nu_x}$ | The mean of the mean minimum salinity | $\nu_x \sim Unif(0,35)$ |
| $\boldsymbol{\nu_y}$ | The mean of the mean maximum salinity | $\nu_y \sim Unif(0,35)$ |

In this BN model, the Bayes' theorem is applied to update our prior distribution with the data. As a result, our posterior distributions denote our updated knowledge about how many species could be introduced in each route. We estimated the posteriors for the parameters using Markov Chain Monte Carlo (MCMC) sampling with the JAGS -software version 4.3.0 (Plummer 2003). We ran the MCMC simulations for 500,000 iterations in three MCMC chains using thinning of 100. The first 200,000 iterations were treated as a burn-in block, thus leaving 9000 samples in the analysis. Convergence was assessed by visual inspection of the chains.

Virta, L., Soininen, J., & Norkko, A. (2020). Diversity and distribution across a large environmental and spatial gradient: Evaluating the taxonomic and functional turnover, transitions and environmental drivers of benthic diatom communities. Global Ecology and Biogeography, 29(12), 2214-2228.

## Appendix C

Table C1. The variable names, the descriptions, the states and the node types. (SWBaltic=southwestern Baltic; IWC=in-water cleaning; NSTM=US Navy fouling rating scale; WSA=wetted surface area; NIS=non-indigenous species)

| Variable name | Description and states | Node type |
|---|---|---|
| Ship type | Different ships: bulker, tanker, cargo, container, passenger, RoRo | Decision |
| Theoretical fuel consumption kg/h | This depends on the ship's unique characteristics. Interval states ranging from 1000 to 5000 kg/h | Decision |
| Fuel type | A ship can use light or heavy fuel: light fuel oil; heavy fuel oil | Decision |
| Annual shipping hours | The hours the ship operates annually:<br>1000-2000; 2000-3000; 3000-4000; 4000-5000; 5000-6000; 6000-7000; 7000-8000; 8000-8760 | Decision |
| Routes | Routes inside the Baltic Sea and routes to and from the Baltic Sea: North Sea-SWBaltic; North Sea-Gulf of Finland; North Sea-Gulf of Bothnia; SWBaltic-Baltic Proper; SWBaltic-Gulf of Riga; Gulf of Finland-Baltic proper; inside the Gulf of Finland; SWBaltic-Baltic Proper; SWBaltic-Baltic Proper; Inside the SWBaltic | Decision |
| Time since coating (years) | The time since the last dry-docking: 0; 1; 2; 3; 4 | Decision |
| Coating type | Different coating types used for hull coating: hard coating; biocidal coating; fouling release coating | Decision |
| In-water cleaning times/growing season | IWC is performed during the growing season (April-September) e.g once-twice in a month: 0; 2; 6; 12 | Decision |
| In-water cleaning method (in the past) | Soft method can be used for soft (biofouling level 0-30 NSTM) but hard methods are needed for hard biofouling (40-100 NSTM): soft brushes; hard brushes | Decision |
| Off hire costs (days) | Off hire costs related to the in-water cleaning if the ship needs to be off hired during the in-water cleaning: none; 1 day; 2 days | Decision |
| In-water cleaning and collecting in the destination port | in-water cleaning releases biofouling material which is collected or not: IWC + no collecting; IWC + collecting; no IWC | Decision |
| Biofouling pressure NSTM average | The amount of organisms attached to the ship hull causes the biofouling level of the hull. Here the average level (%) of biofouling is considered: 0-10; 10-20; 20-30; 30-40; 40-50; 60-70; 70-80; 80-90; 90-100 | Random |
| Fuel consumption real kg/year | The real amount of fuel used by a ship annually: 1000-2000; 2000-3000; 3000-4000; 4000-5000; 5000-6000; 6000-7000 | Random |
| Air emissions ($CO_2$) kg/h | The $CO_2$ emissions depend on the fuel type and fuel consumption | Random |
| Wetted surface area (WSA) | The whole size of the area of the immersed ship hull: interval 0-1.04 $km^2$ | Random |
| WSA without niche | The size of the area of the immersed ship hull without niche areas: interval between 0-100 $hm^2$ | Random |
| Niche areas $hm^2$ | The size of the areas in the hull favorable for attachment: range between: 0-30 $hm^2$ | Random |
| Biofouling pressure NSTM maximum | The amount of organisms attached to the ship hull causes the biofouling level of the hull. Here the maximum level (%) of biofouling is considered: 0-10; 10-20; 20-30; 30-40; 40-50; 60-70; 70-80; 80-90; 90-100 | Random |
| Fouling type | The length and hardness of the hull biofouling: soft fouling (0-30 NSTM) ; hard fouling (40-100 NSTM) | Random |
| NIS value | The actual number of non-indigenous species (NIS) able to be transported from the source port to the destination port: numbered between 1-53 | Random |
| Potential risk in niche areas | The risk depends on the biofouling pressure, NIS value and the size of niche area: interval between 0-45000 | Random |
| Potential risk in WSA (without niche areas) | The risk depends on the biofouling pressure, NIS value and the size of WSA without niche areas: interval between 0-55000 | Random |
| Copper emissions µg/cm$^2$/day | The average amount of copper the coating is releasing to the water (µg/cm$^2$/day): 0; 7 | Random |
| Ecotoxicological pressure kg/year | The copper release to water depends on the WSA, coating type, IWC times/growing season and IWC methods: interval between 0-45000 | Random |
| Sediment copper concentration | The current copper concentration in the sediment is considered here only as low; high | Random |
| Fuel costs €/year | The annual ship fuel costs depend on the fuel amount and the fuel type | Utility (costs) |
| Fuel costs €/h | The ship fuel costs per hour depend on the fuel amount and the fuel type | Utility (costs) |
| $CO_2$ emissions kg/year | The amount of $CO_2$ released from the fuel consumption annually | Utility (environmental impacts) |
| $CO_2$ emissions kg/h | The amount of $CO_2$ released from the fuel consumption per hour | Utility (environmental impacts) |
| In-water cleaning costs €/year | The IWC costs depend on the ship type, size, method used and the cleaning times | Utility (costs) |
| Coating costs €/year | The costs to coat the ship hull depends on the coating type and the ship size | Utility (costs) |
| NIS introduction risk/arrival | The conditional risk of potential non-indigenous species introduction per arrival to the port | Utility (environmental impacts) |
| Ecotoxicological risk kg/year | The annual copper emissions released to water from the coating | Utility (environmental impacts) |
| Sediment eco-toxicological risk | The risk is calculated based on the current sediment copper concentration in a certain area and the copper released to the water from the biocidal coating | Utility (environmental impacts) |

# The user manual for the decision support tool for ship biofouling management in the Baltic Sea

## Foreword

This document provides a user manual for the decision support tool for ship biofouling management in the Baltic Sea. The tool was developed in the project COMPLETE - Completing management options in the Baltic Sea region to reduce risk of invasive species introduction by shipping [#R069] (https://www.balticcomplete.com/). The COMPLETE project is co-financed by the European Union's funding Programme Interreg Baltic Sea Region (European Regional Development Fund).

Biofouling, the accumulation of organism e.g. on ship hull, increases the fuel consumption but also the risk for harmful non-indigenous species (NIS) introduction which may have severe impacts in the sensitive Baltic Sea. The shipping industry uses different methods to decrease the biofouling level but some of the methods are harmful for the environment. This decision support tool can be used for studying the optimal biofouling management case-specifically in the Baltic Sea. The tool provides a holistic view on biofouling-associated risks and increases the systemic understanding of the complex problem with multiple social, ecological and economic aspects: eco-toxicological risk related to use of biocidal coating, NIS introduction risk related to biofouling, CO2 emissions related to fuel consumption and costs related to fuel consumption and the biofouling management (coating and in-water cleaning). In addition, the tool can support the process in making the biofouling management plan. When developing biofouling regulations, the possibility to study the problem with a tool providing a comprehensive view can benefit the authorities as well.

The tool (structure and probability distributions) is based on the data collected in the project from different sources: literature, experts, stakeholders, AquaNIS and on-board measurements. The tool is an influence diagram (ID) model, more specific a Bayesian network (BN) model and can be opened with a BN software that supports utility and decision nodes (see softwares available: https://www.cs.ubc.ca/~murphyk/Software/bnsoft.html). The tool is developed using the software Hugin (https://www.hugin.com/) and this manual offers instructions for conducting the analyses with that software. This tool is free for use (See Appendix E).

# 1. The structure of the decision support tool

## 1.1 Structure of the model

The graphical tool consists of three types of variables: 11 decision (rectangles), 14 random (ovals) and 9 utility (diamonds) variables and links (arrows) representing their conditional dependencies (Fig. D1).

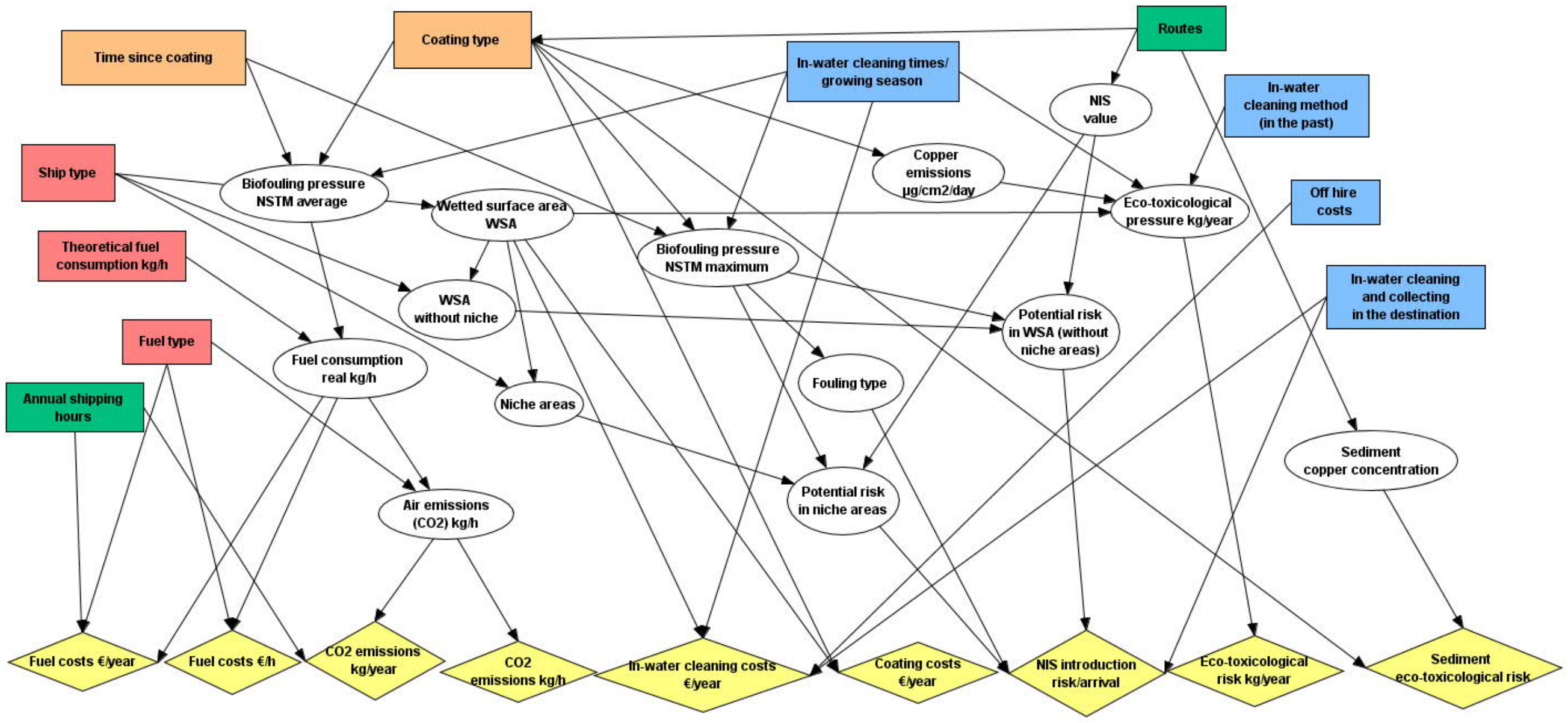

Figure D1. The structure of the decision support tool.

The variables in the model are from now on called nodes. The nodes with incoming arrows are dependent variables, of which state is conditional on other variables (see Fig. D2). On the contrary, the state of the nodes without any incoming arrows is unconditional, i.e. *independent of any other variable*. For the detailed information considering the variables, see the main article and Appendix C.

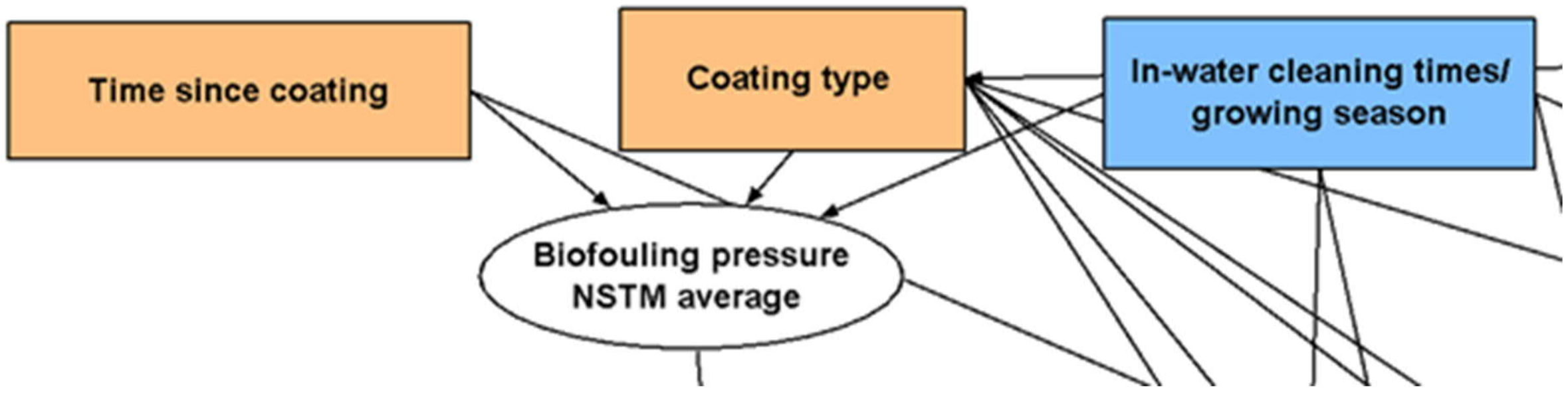

Figure D2. The decision node "Time since coating" is an independent variable, unaffected by any other variable. On the contrary, the state of the random node "Biofouling pressure NSTM average" is a dependent variable, affected by the state of three other variables, indicated by the incoming arrows.

## Random nodes

The random (oval-shaped) nodes represent the key elements of the system through which the decisions made together impact on the state of the utility nodes. In this model, the random nodes are dependent on decision nodes directly or through other random nodes. Thus, each random node has several possible states, the probabilities of which depend on the states of the preceding random and decision nodes.

In the edit mode (the Hugin software program opens in the edit mode) node-specific information can be seen by *double-clicking* the node. In Fig. D3, the appearing of the "information box" of the random node *Air emissions CO2 kg/h* can be seen. *Click the "States" -sheet* to see the (possible) states of the node.

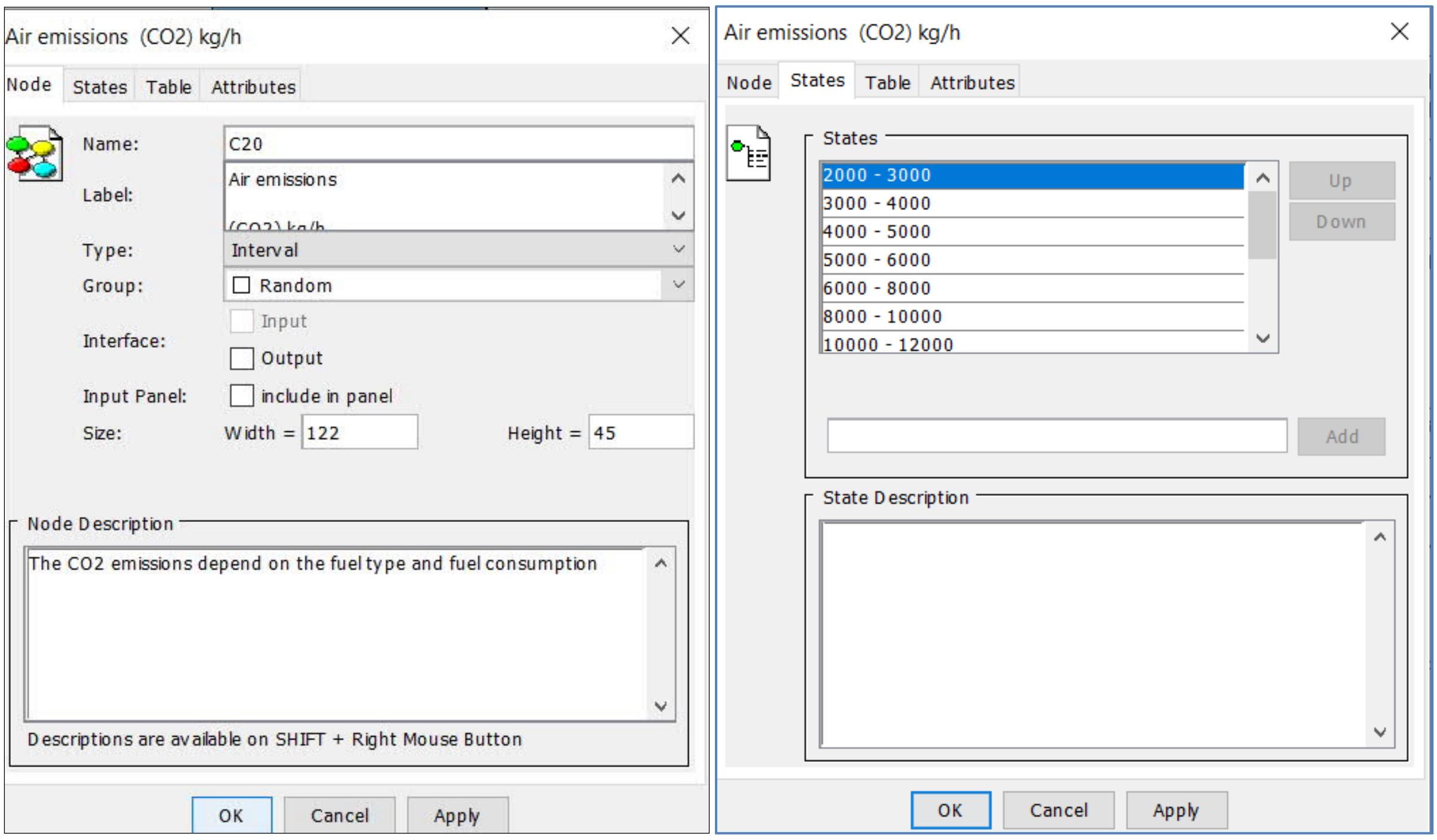

Figure D3. Node-related information, which can be seen by *double-clicking* the node.

In the edit mode, the probability tables of the random nodes can be viewed by activating a node by clicking on it with left click. Then open the node menu by right clicking and choose *"Open tables"*. The probability table of the node should now open. When the random node has incoming links (i.e. dependent variable), the tables behind the nodes are called conditional probability tables. These tables show the probability distribution for each combination of the alternative states of their nodes from which the links come (see Figures D2 and D4). In Fig. D4, the conditional probability table of the random node *Biofouling pressure NSTM average* shows how the decision nodes *Time since last coating*, *Coating type* and *In-water cleaning times/growing season* have an impact on the level of the biofouling.

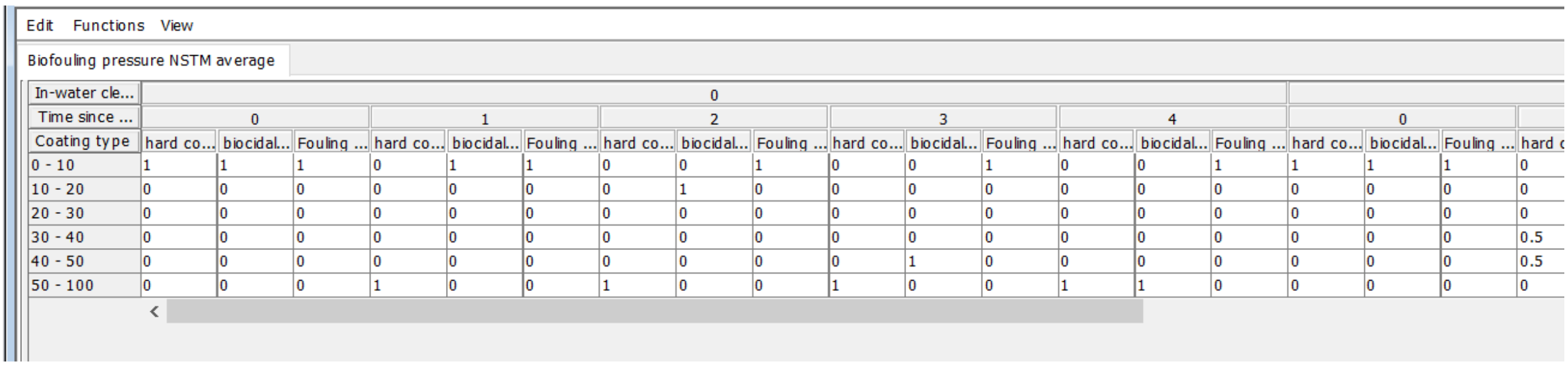

Edit Functions View

Biofouling pressure NSTM average

| In-water cle... | 0 | | | | | | | | | | | | | | | | | | |
|---|---|---|---|---|---|---|---|---|---|---|---|---|---|---|---|---|---|---|---|
| Time since ... | 0 | | | 1 | | | 2 | | | 3 | | | 4 | | | 0 | | | |
| Coating type | hard co... | biocidal... | Fouling ... | hard co... | biocidal... | Fouling ... | hard co... | biocidal... | Fouling ... | hard co... | biocidal... | Fouling ... | hard co... | biocidal... | Fouling ... | hard co... | biocidal... | Fouling ... | hard c |
| 0 - 10 | 1 | 1 | 1 | 0 | 1 | 1 | 0 | 0 | 1 | 0 | 0 | 1 | 0 | 0 | 1 | 1 | 1 | 1 | 0 |
| 10 - 20 | 0 | 0 | 0 | 0 | 0 | 0 | 0 | 1 | 0 | 0 | 0 | 0 | 0 | 0 | 0 | 0 | 0 | 0 | 0 |
| 20 - 30 | 0 | 0 | 0 | 0 | 0 | 0 | 0 | 0 | 0 | 0 | 0 | 0 | 0 | 0 | 0 | 0 | 0 | 0 | 0 |
| 30 - 40 | 0 | 0 | 0 | 0 | 0 | 0 | 0 | 0 | 0 | 0 | 0 | 0 | 0 | 0 | 0 | 0 | 0 | 0 | 0.5 |
| 40 - 50 | 0 | 0 | 0 | 0 | 0 | 0 | 0 | 0 | 0 | 0 | 1 | 0 | 0 | 0 | 0 | 0 | 0 | 0 | 0.5 |
| 50 - 100 | 0 | 0 | 0 | 1 | 0 | 0 | 1 | 0 | 0 | 1 | 0 | 0 | 1 | 1 | 0 | 0 | 0 | 0 | 0 |

Figure D4. Illustration of the conditional probability table of the random node *Biofouling pressure NSTM average* in the edit mode.

Figure D4 shows that even if the in-water cleaning times during the growing season is zero, *Biofouling pressure NSTM average* is in the lowest state (the state of the node is between 0 and 10) if the time since coating is also zero and coating type is either hard, biocidal or fouling release coating. In the run mode (press the run mode tool button with lightning figure to run the model), probability distribution of the random node *Biofouling pressure NSTM average* show the states for the biofouling pressure that are dependent on the states of preceding decision nodes (Fig. D5).

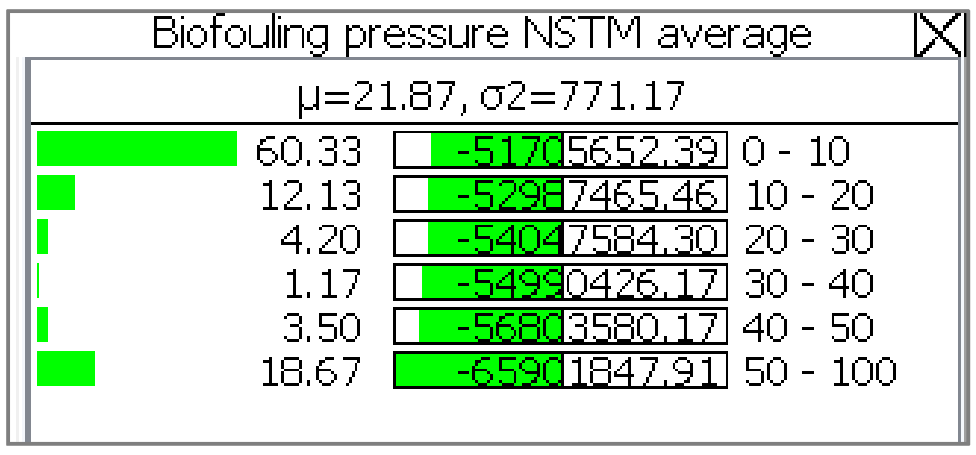

Figure D5. Illustration of the probability distribution of the random node *Biofouling pressure NSTM average* in the run mode.

## Decision and utility nodes

The decision (rectangle-shaped) nodes enable the case specific biofouling management analysis. There are four categories of decision nodes in the model, illustrated with different colors: three ship-related (red), two operational decisions (green), two coating-related (orange) and three in-water cleaning -related (blue). Each decision node has a different number of alternative states, of which the end-user is intended to select the suitable one (e.g. to match the profile of the ship to be analyzed) and hence test the effects of different combinations of management options.

The decision node *Routes* has ten different states (1-10) but each route has two directions: from the port A to the port B and the other way round (Table D1). This way the user can study whether the gains and losses related to the biofouling management differ depending on the direction. The direction of the route affects especially on the risk related in-water cleaning and the eco-toxicity. The NIS introduction risk and sediment copper concentration varies in different areas. Notice that fouling release coating can be used only in ice free areas year-round such as the southern Baltic Sea. For further information considering the decision nodes see the main article and Appendix C.

Table D1. Different states (1-10) of the decision node *Routes.* Each route has two directions, i.e. from the port A to the port B and the other way round.

| ROUTE | DEPARTURE AREA | ARRIVAL AREA |
|---|---|---|
| 1A | The North Sea (e.g. Rotterdam) | The Southwestern Baltic (e.g. Trelleborg) |
| 1B | The Southwestern Baltic (e.g. Trelleborg) | The North Sea (e.g. Rotterdam) |
| 2A | The North Sea (e.g. Rotterdam) | The Gulf of Finland (e.g. Saint Petersburg) |
| 2B | The Gulf of Finland (e.g. Saint Petersburg) | The North Sea (e.g. Rotterdam) |
| 3A | The North Sea (e.g. Rotterdam) | The Gulf of Bothnia (e.g. Oulu) |
| 3B | The Gulf of Bothnia (e.g. Oulu) | The North Sea (e.g. Rotterdam) |
| 4A | The Southwestern Baltic (e.g. Copenhagen) | The Baltic Proper (e.g. Stockholm) |
| 4B | The Baltic Proper (e.g. Stockholm) | The Southwestern Baltic (e.g. Copenhagen) |
| 5A | The Southwestern Baltic (e.g. Århus) | The Gulf of Riga (e.g. Riga) |
| 5B | The Gulf of Riga (e.g. Riga) | The Southwestern Baltic (e.g. Århus) |
| 6A | The Gulf of Finland (e.g. Helsinki) | The Baltic proper (e.g. Stockholm) |
| 6B | The Baltic proper (e.g. Stockholm) | The Gulf of Finland (e.g. Helsinki) |
| 7A | The Gulf of Finland (e.g. Helsinki) | The Gulf of Finland (e.g. Tallinn) |
| 7B | The Gulf of Finland (e.g. Tallinn) | The Gulf of Finland (e.g. Helsinki) |
| 8A | The Southwestern Baltic (e.g. Rostock) | The Baltic Proper (e.g. Gdansk) |
| 8B | The Baltic Proper (e.g. Gdansk) | The southwestern Baltic(e.g. Rostock) |
| 9A | The Southwestern Baltic (e.g. Kiel) | The Baltic Proper (e.g. Riga) |
| 9B | The Baltic Proper (e.g. Riga) | The Southwestern Baltic (e.g. Kiel) |
| 10A | The Southwestern Baltic (e.g. Malmo) | The Southwestern Baltic (e.g. Rostock) |
| 10B | The Southwestern Baltic (e.g. Rostock) | The Southwestern Baltic (e.g. Malmo) |

The utility (diamond-shaped) nodes play a key role when the aim is to compare alternative decisions, as they present the expected gains and losses of each scenario. Even though the units of the utility nodes are not commensurable, gains and losses of the different scenarios can be compared. All the utilities are calculated per annum, except the utility nodes *NIS introduction risk/arrival* and *Sediment eco-toxicological risk*. The *NIS introduction risk/arrival* presents the risk of each arrival to the port and thus depends on the annual arrivals. The *sediment eco-toxicological risk* depends on the coating type used and the existing sediment copper concentration mg/kg; biocidal coating with in-water cleaning increases the risk while hard coating or fouling release coating does not.

The aim of the decision support tool is to illustrate case-specific biofouling management strategies, which minimize the costs and environmental harms. The user is intended to set the state of decision nodes, affecting the state of the different random nodes and finally to the gains and losses in the utility nodes. The utility nodes cover different costs related to biofouling and its management and fuel consumption, carbon dioxide emissions resulting from fuel consumption as well as non-indigenous species introduction risk and eco-toxicological risk due to biocidal coating.

## 2. Conducting a case-specific analysis

A thorough instruction of running an analysis of the decision support tool by using the Hugin software:

1. When the Hugin software is opened, it opens in the edit mode. In this mode, you can make some modifications to the model structure. For instance, you can modify the prices related to the coating (utility node *Coating cost €/year*) and fuels (utility node *Fuel costs €/h*). By opening the probability table of the node, you can change the values in the Expression line. After possible modifications, it is possible to run the model by clicking the lightning icon.

2. In the run mode (Fig. D6), the network window is divided into the node list panel (on the left) and the network panel (on the right). To conduct a case-specific analysis, select a certain state for a decision node by activating the node by *clicking on it*, then *right click* and choose "Show Monitor Window(s)". The monitor window for a decision node shows up. Then *double-click* the state you want to choose from the monitor window. The selected state "turns to red". Repeat the steps for each decision node, of which status you want to lock. When the wanted states are locked, you can examine the impacts of different scenarios by reading the expected utilities. To see the values for different utility nodes, you can open the monitor windows of the nodes. Alternatively, you can activate the utility node in the node list panel by *double-clicking* the labels of the utility node. To unlock the state, *double-click* the node again. All the states return to green. Furthermore, by *clicking* initialize network *(arrow-ended circle icon)*, all decisions return to green, unlocked state.

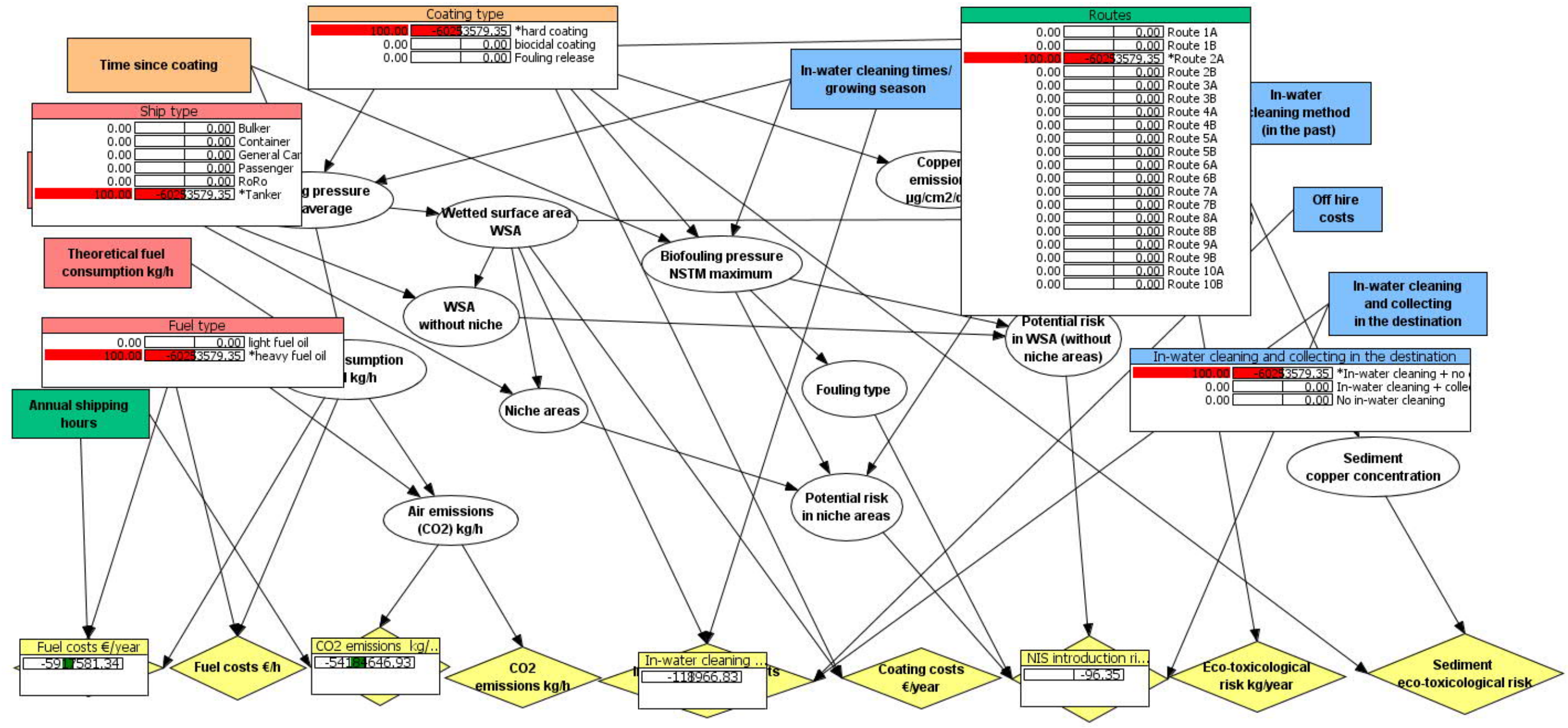

Figure D6. Illustration of the run mode.

3. At this state, you can run the model with your own scenario(s) (by locking the wanted states of decision nodes). NB: If you decide not to lock the states, those decision nodes have a uniform distribution meaning each states have an equal probability. This also influences the results, and you should take it into account when making interpretations.

   When you run the model, you can leave the *Time since coating* and *In-water cleaning method (in the past)* unlocked, but for the results to be reasonable, the other decision nodes are recommended to be locked. In addition, it is recommended to lock the random node *Wetted surface area* after locking the decision node *Ship type.* This allows you to lock to a state that corresponds more correctly to the ship size. Notice that if you lock the state No in-water cleaning in the decision node I*n-water cleaning and collecting in destination port,* you must also lock the state 0 in the node *In-water cleaning times/ growing season*.

   When interpreting the expected utilities, it is important to note that the values of the utility nodes are negative indicating undesirable outcomes. Since the negative values are only used to illustrate the undesirability of the outcome, ranking between the values can be made similarly than to positive numbers. For instance, the utility node *NIS introduction risk/arrival* with value -1.27 represents a higher risk than the risk with value -0.60. The further the value is from zero, the more undesirable is the outcome. Finally, the units between the utility nodes vary. For instance, all costs are measured in terms of euros, but Ecotoxicological risk is measured as kg/year and $CO_2$ emissions as kg/year, while NIS-introduction risk is a pure number. Thus, except for the costs, the utilities are not comparable. However, the results of the utilities in different scenarios can be compared.

For instance, in the Figure D6, we have selected the tanker using heavy oil and hard coating in the route between the North Sea and the Gulf of Finland (the route 2A) for the analysis. In this example, you can study how the different in-water cleaning methods in the destination port impact on the expected utilities *NIS introduction risk/arrival* and *In-water cleaning cost €/year*. For instance, if the state in-water cleaning and collecting (decision node *In-water cleaning and collecting in the destination*) is locked, the expected utilities for the *NIS introduction risk/arrival* and *In-water cleaning cost €/year* are –27.25 and –128450.25, respectively. However, if we would lock the state in-water cleaning and no collecting, we could see that the expected utility for the *NIS introduction risk/arrival* decreases to -96.35 but the *In-water cleaning cost €/year* increases slightly to –118966.83.

# 3. Final remarks

This decision support tool is meant to support the holistic understanding about the complex biofouling management issue in the Baltic Sea region as well as to help to recognize the potential management options and their consequences. The tool allows the user to study visually and transparently the main antifouling methods (i.e. coating types and in-water cleaning methods) jointly in a way that the user can compare the risks and costs related to monetary costs (biofouling and fuel consumption), $CO_2$ emissions, ecotoxicological risk (copper load) and NIS introduction risk of each method.

When using the tool, the user should consider the assumptions of the model that might have an impact on the results. The main assumptions of the model:

- ➔ Modeled situation contains a single voyage (which can be repeated several times under selected conditions) with certain biofouling level and environmental risk due to ship's previous management methods
- ➔ The model works best for the ships sailing on fixed routes
- ➔ Operations under ice conditions or other seasonal variations are excluded
- ➔ Biofouling occurs only during the growing season and possible ice scraping the hull and releasing the biofouling material is not considered
- ➔ The hard IWC method damaging the hull is not considered and the degree of wear is based on the age of the coating

- The idle time and its impact on biofouling level and NIS introduction risk is not considered
- The release rate of copper and the biofouling level are assumed to be the same in the whole Baltic Sea
- The hull can have only one coating type
- Only $CO_2$ is taken into account for the emission level and copper for the eco-toxicological risk
- Only salinity is considered as an environmental factor limiting the NIS introductions
- The accumulation of biofouling is assumed to occur evenly on the ship hull
- The bioaccumulation of the released material is not taken into account
- NIS introduction risk is calculated only for the departure port where the possible IWC occurs (the NIS introduction risk occurring during the sail is acknowledged)

# Appendix E

```
//Auto-generated C code from HUGIN 8800 (double precision, table generator: true)
//when function returns, verify that HUGIN objects was successfully created by asserting (h_error_code() == h_error_none)

# include "hugin.h"
# include <string.h>

h_class_collection_t create_class_collection() {
    //define
    h_class_collection_t cc;
    h_class_t cls_COMPLETE_DST_manual;
    h_node_t node_buffer[4];
    h_table_t t;
    h_number_t* d;
    size_t i, j;
    h_model_t model;
    h_expression_t expr;
    h_node_t node_COMPLETE_DST_manual_D4;
    h_node_t node_COMPLETE_DST_manual_U9;
    h_node_t node_COMPLETE_DST_manual_C17;
    h_node_t node_COMPLETE_DST_manual_C16;
    h_node_t node_COMPLETE_DST_manual_C1;
    h_node_t node_COMPLETE_DST_manual_U8;
    h_node_t node_COMPLETE_DST_manual_D2;
    h_node_t node_COMPLETE_DST_manual_C21;
    h_node_t node_COMPLETE_DST_manual_C19;
    h_node_t node_COMPLETE_DST_manual_U7;
    h_node_t node_COMPLETE_DST_manual_C18;
    h_node_t node_COMPLETE_DST_manual_C14;
    h_node_t node_COMPLETE_DST_manual_C13;
    h_node_t node_COMPLETE_DST_manual_Routes;
    h_node_t node_COMPLETE_DST_manual_C10;
    h_node_t node_COMPLETE_DST_manual_C6;
    h_node_t node_COMPLETE_DST_manual_D1;
    h_node_t node_COMPLETE_DST_manual_C9;
    h_node_t node_COMPLETE_DST_manual_U6;
    h_node_t node_COMPLETE_DST_manual_U5;
    h_node_t node_COMPLETE_DST_manual_C3;
    h_node_t node_COMPLETE_DST_manual_C8;
    h_node_t node_COMPLETE_DST_manual_C4;
    h_node_t node_COMPLETE_DST_manual_D7;
    h_node_t node_COMPLETE_DST_manual_D6;
    h_node_t node_COMPLETE_DST_manual_C12;
    h_node_t node_COMPLETE_DST_manual_U4;
    h_node_t node_COMPLETE_DST_manual_C20;
    h_node_t node_COMPLETE_DST_manual_D3;
    h_node_t node_COMPLETE_DST_manual_C11;
    h_node_t node_COMPLETE_DST_manual_U3;
    h_node_t node_COMPLETE_DST_manual_U2;
    h_node_t node_COMPLETE_DST_manual_U1;
    h_node_t node_COMPLETE_DST_manual_C5;

    //string constants
    const char* s[136] = {"D4", "U9", "C17", "C16", "C1", "U8", "D2", "C21", "C19", "U7", "C18", "C14", "C13", "Routes", "C10", "C6", "D1", "C9", "U6", "U5", "C3", "C8", "C4", "D7", "D6",
"C12", "U4", "C20", "D3", "C11", "U3", "U2", "U1", "C5", "None", "1 day", "2 days", "hard coating", "biocidal coating", "Fouling release", "Soft fouling", "Hard fouling", "Route 1A", "Route
1B", "Route 2A", "Route 2B", "Route 3A", "Route 3B", "Route 4A", "Route 4B", "Route 5A", "Route 5B", "Route 6A", "Route 6B", "Route 7A", "Route 7B", "Route 8A", "Route 8B", "Route
9A", "Route 9B", "Route 10A", "Route 10B", "In-water cleaning + no collecting", "In-water cleaning + collecting", "No in-water cleaning", "Soft brushes", "Hard brushes", "low", "high",
"light fuel oil", "heavy fuel oil", "Bulker", "Container", "General Cargo", "Passenger", "RoRo", "Tanker", "-C20*D2", "-C11*0.5*D2", "-C11*0.35*D2", "0", "365*7*(C8*1E+010)/1E+009",
"(2*7*12*(C8*1E+010)+(365-2*7)*7*(C8*1E+010))/1E+009", "(6*7*12*(C8*1E+010)+(365-6*7)*7*(C8*1E+010))/1E+009", "(12*7*12*(C8*1E+010)+(365-
12*7)*7*(C8*1E+010))/1E+009", "(2*7*25*(C8*1E+010)+(365-2*7)*7*(C8*1E+010))/1E+009", "(6*7*25*(C8*1E+010)+(365-6*7)*7*(C8*1E+010))/1E+009",
"(12*7*25*(C8*1E+010)+(365-12*7)*7*(C8*1E+010))/1E+009", "(C8-C8*0.07)*100", "(C8-C8*0.09)*100", "(C8-C8*0.27)*100", "(C8-C8*0.08)*100", "C17*2.6*C10*C3/10",
"C17*C19*C3/10", "C8*0.07*100", "C8*0.09*100", "C8*0.27*100", "C8*0.15*100", "C8*0.08*100", "-C8*1E+006*32/5", "-C8*1E+006*22/5", "-C8*1E+006*52/5", "-
C8*1E+006*0.4*3*2", "-C8*1E+006*0.4*3*6", "-C8*1E+006*0.4*3*12", "-(C8*1E+006*0.4*3*2+2*20000)", "-(C8*1E+006*0.4*3*6+6*20000)", "-(C8*1E+006*0.4*3*12+12*20000)", "-
(C8*1E+006*0.4*3*2+2*40000)", "-(C8*1E+006*0.4*3*6+6*40000)", "-(C8*1E+006*0.4*3*12+12*40000)", "-C8*1E+006*0.4*3*2*1.5", "-C8*1E+006*0.4*3*6*1.5", "-
C8*1E+006*0.4*3*12*1.5", "-(C8*1E+006*0.4*3*2*1.5+2*20000)", "-(C8*1E+006*0.4*3*6*1.5+6*20000)", "-(C8*1E+006*0.4*3*12*1.5+12*20000)", "-
(C8*1E+006*0.4*3*2*1.5+2*40000)", "-(C8*1E+006*0.4*3*6*1.5+6*40000)", "-(C8*1E+006*0.4*3*12*1.5+12*40000)", "C11*2.7", "C11*3.2", "D7*1.02", "D7*1.04", "D7*1.06",
"D7*1.08", "D7*1.12", "D7*1.3", "-(C13*0.9*0.9+C14*0.05)", "-(C13*0.9*0.05*0.9+C14*0.05)", "-(C13+C14)*0.05", "-(C13*0.6*0.08+C14*0.05)", "-(C13*0.6*0.05*0.08+C14*0.05)", "-
C11*0.5", "-C11*0.35", "COMPLETE_DST_manual"};

    //some numbers
    static h_number_t arr_0_0[] = {0.0,0.0,0.0,0.0,0.0,0.0,0.0,0.0,0.0,0.0,0.0,0.0,0.0,0.0,0.0,0.0,0.0,0.0,0.0,0.0,0.0,0.0,0.0,0.0};
    static h_number_t arr_0_5[] = {0.5,0.5};
    static h_number_t arr_0_3[] = {0.3,0.3,0.3,0.3,0.3,0.3};
    static h_number_t arr_3_92517E__5[] = {3.92517E-5,3.92517E-5,3.92517E-5,3.92517E-5,3.92517E-5,3.92517E-5,3.92517E-5};
    static h_number_t arr_5_88568E__5[] = {5.88568E-5,5.88568E-5,5.88568E-5,5.88568E-5,5.88568E-5,5.88568E-5,5.88568E-5};
    static h_number_t arr_8_82073E__5[] = {8.82073E-5,8.82073E-5,8.82073E-5,8.82073E-5,8.82073E-5,8.82073E-5,8.82073E-5};
    static h_number_t arr_2_14085E__4[] = {2.14085E-4,2.14085E-4};
    static h_number_t arr_1_58873E__4[] = {1.58873E-4,1.58873E-4,1.58873E-4,1.58873E-4,1.58873E-4,1.58873E-4,1.58873E-4};
    static h_number_t arr_1_65322E__4[] = {1.65322E-4,1.65322E-4,1.65322E-4,1.65322E-4,1.65322E-4,1.65322E-4,1.65322E-4};
```

```
if ((cc = h_new_class_collection()) == 0)
    return cc;

//initialize class COMPLETE_DST_manual
if ((cls_COMPLETE_DST_manual = h_cc_new_class(cc)) == 0 ||
    h_class_set_name(cls_COMPLETE_DST_manual, (h_string_t)s[135]) != 0)
    return cc;

if ((node_COMPLETE_DST_manual_D4 = h_class_new_node(cls_COMPLETE_DST_manual, h_category_decision, h_kind_discrete)) == 0 ||
    h_node_set_name(node_COMPLETE_DST_manual_D4, (h_string_t)s[0]) != 0 ||
    h_node_set_number_of_states(node_COMPLETE_DST_manual_D4, 3) != 0)
    return cc;
{
    static size_t label[] = {34, 35, 36};
    for (i = 0; i < 3; ++i) {
        if (h_node_set_state_label(node_COMPLETE_DST_manual_D4, i, (h_string_t)s[label[i]]) != 0)
            return cc;
    }
}

if ((node_COMPLETE_DST_manual_U9 = h_class_new_node(cls_COMPLETE_DST_manual, h_category_utility, h_kind_other)) == 0 ||
    h_node_set_name(node_COMPLETE_DST_manual_U9, (h_string_t)s[1]) != 0)
    return cc;

if ((node_COMPLETE_DST_manual_C17 = h_class_new_node(cls_COMPLETE_DST_manual, h_category_chance, h_kind_discrete)) == 0 ||
    h_node_set_name(node_COMPLETE_DST_manual_C17, (h_string_t)s[2]) != 0 ||
    h_node_set_number_of_states(node_COMPLETE_DST_manual_C17, 6) != 0 ||
    h_node_set_subtype(node_COMPLETE_DST_manual_C17, h_subtype_interval) != 0)
    return cc;
{
    static h_number_t value[] = {0.0, 10.0, 20.0, 30.0, 40.0, 50.0, 100.0};
    for (i = 0; i <= 6; ++i) {
        if (h_node_set_state_value(node_COMPLETE_DST_manual_C17, i, value[i]) != 0)
            return cc;
    }
}

if ((node_COMPLETE_DST_manual_C16 = h_class_new_node(cls_COMPLETE_DST_manual, h_category_decision, h_kind_discrete)) == 0 ||
    h_node_set_name(node_COMPLETE_DST_manual_C16, (h_string_t)s[3]) != 0 ||
    h_node_set_number_of_states(node_COMPLETE_DST_manual_C16, 3) != 0)
    return cc;
{
    static size_t label[] = {37, 38, 39};
    for (i = 0; i < 3; ++i) {
        if (h_node_set_state_label(node_COMPLETE_DST_manual_C16, i, (h_string_t)s[label[i]]) != 0)
            return cc;
    }
}

if ((node_COMPLETE_DST_manual_C1 = h_class_new_node(cls_COMPLETE_DST_manual, h_category_chance, h_kind_discrete)) == 0 ||
    h_node_set_name(node_COMPLETE_DST_manual_C1, (h_string_t)s[4]) != 0 ||
    h_node_set_number_of_states(node_COMPLETE_DST_manual_C1, 6) != 0 ||
    h_node_set_subtype(node_COMPLETE_DST_manual_C1, h_subtype_interval) != 0)
    return cc;
{
    for (i = 0; i <= 6; ++i) {
        if (h_node_set_state_value(node_COMPLETE_DST_manual_C1, i, h_node_get_state_value(node_COMPLETE_DST_manual_C17, i)) != 0)
            return cc;
    }
}

if ((node_COMPLETE_DST_manual_U8 = h_class_new_node(cls_COMPLETE_DST_manual, h_category_utility, h_kind_other)) == 0 ||
    h_node_set_name(node_COMPLETE_DST_manual_U8, (h_string_t)s[5]) != 0)
    return cc;

if ((node_COMPLETE_DST_manual_D2 = h_class_new_node(cls_COMPLETE_DST_manual, h_category_decision, h_kind_discrete)) == 0 ||
    h_node_set_name(node_COMPLETE_DST_manual_D2, (h_string_t)s[6]) != 0 ||
    h_node_set_number_of_states(node_COMPLETE_DST_manual_D2, 8) != 0 ||
    h_node_set_subtype(node_COMPLETE_DST_manual_D2, h_subtype_interval) != 0)
    return cc;
{
    static h_number_t value[] = {1000.0, 2000.0, 3000.0, 4000.0, 5000.0, 6000.0, 7000.0, 8000.0, 8760.0};
    for (i = 0; i <= 8; ++i) {
        if (h_node_set_state_value(node_COMPLETE_DST_manual_D2, i, value[i]) != 0)
            return cc;
    }
}

if ((node_COMPLETE_DST_manual_C21 = h_class_new_node(cls_COMPLETE_DST_manual, h_category_chance, h_kind_discrete)) == 0 ||
    h_node_set_name(node_COMPLETE_DST_manual_C21, (h_string_t)s[7]) != 0 ||
    h_node_set_number_of_states(node_COMPLETE_DST_manual_C21, 15) != 0 ||
    h_node_set_subtype(node_COMPLETE_DST_manual_C21, h_subtype_interval) != 0)
```

```
    return cc;
{
    static h_number_t value[] = {0.0, 0.0, 25.0, 50.0, 100.0, 500.0, 1000.0, 5000.0, 10000.0, 15000.0, 20000.0, 25000.0, 30000.0, 35000.0, 40000.0, 45000.0};
    for (i = 0; i <= 15; ++i) {
        if (h_node_set_state_value(node_COMPLETE_DST_manual_C21, i, value[i]) != 0)
            return cc;
    }
}

if ((node_COMPLETE_DST_manual_C19 = h_class_new_node(cls_COMPLETE_DST_manual, h_category_chance, h_kind_discrete)) == 0 ||
    h_node_set_name(node_COMPLETE_DST_manual_C19, (h_string_t)s[8]) != 0 ||
    h_node_set_number_of_states(node_COMPLETE_DST_manual_C19, 20) != 0 ||
    h_node_set_subtype(node_COMPLETE_DST_manual_C19, h_subtype_interval) != 0)
    return cc;
{
    static h_number_t value[] = {0.0, 5.0, 10.0, 15.0, 20.0, 25.0, 30.0, 35.0, 40.0, 45.0, 50.0, 55.0, 60.0, 65.0, 70.0, 75.0, 80.0, 85.0, 90.0, 95.0, 100.0};
    for (i = 0; i <= 20; ++i) {
        if (h_node_set_state_value(node_COMPLETE_DST_manual_C19, i, value[i]) != 0)
            return cc;
    }
}

if ((node_COMPLETE_DST_manual_U7 = h_class_new_node(cls_COMPLETE_DST_manual, h_category_utility, h_kind_other)) == 0 ||
    h_node_set_name(node_COMPLETE_DST_manual_U7, (h_string_t)s[9]) != 0)
    return cc;

if ((node_COMPLETE_DST_manual_C18 = h_class_new_node(cls_COMPLETE_DST_manual, h_category_chance, h_kind_discrete)) == 0 ||
    h_node_set_name(node_COMPLETE_DST_manual_C18, (h_string_t)s[10]) != 0 ||
    h_node_set_number_of_states(node_COMPLETE_DST_manual_C18, 2) != 0)
    return cc;
{
    static size_t label[] = {40, 41};
    for (i = 0; i < 2; ++i) {
        if (h_node_set_state_label(node_COMPLETE_DST_manual_C18, i, (h_string_t)s[label[i]]) != 0)
            return cc;
    }
}

if ((node_COMPLETE_DST_manual_C14 = h_class_new_node(cls_COMPLETE_DST_manual, h_category_chance, h_kind_discrete)) == 0 ||
    h_node_set_name(node_COMPLETE_DST_manual_C14, (h_string_t)s[11]) != 0 ||
    h_node_set_number_of_states(node_COMPLETE_DST_manual_C14, 14) != 0 ||
    h_node_set_subtype(node_COMPLETE_DST_manual_C14, h_subtype_interval) != 0)
    return cc;
{
    static h_number_t value[] = {0.0, 25.0, 50.0, 100.0, 500.0, 1000.0, 5000.0, 10000.0, 15000.0, 20000.0, 25000.0, 30000.0, 35000.0, 40000.0, 45000.0};
    for (i = 0; i <= 14; ++i) {
        if (h_node_set_state_value(node_COMPLETE_DST_manual_C14, i, value[i]) != 0)
            return cc;
    }
}

if ((node_COMPLETE_DST_manual_C13 = h_class_new_node(cls_COMPLETE_DST_manual, h_category_chance, h_kind_discrete)) == 0 ||
    h_node_set_name(node_COMPLETE_DST_manual_C13, (h_string_t)s[12]) != 0 ||
    h_node_set_number_of_states(node_COMPLETE_DST_manual_C13, 16) != 0 ||
    h_node_set_subtype(node_COMPLETE_DST_manual_C13, h_subtype_interval) != 0)
    return cc;
{
    static h_number_t value[] = {0.0, 25.0, 50.0, 100.0, 500.0, 1000.0, 5000.0, 10000.0, 15000.0, 20000.0, 25000.0, 30000.0, 35000.0, 40000.0, 45000.0, 50000.0, 55000.0};
    for (i = 0; i <= 16; ++i) {
        if (h_node_set_state_value(node_COMPLETE_DST_manual_C13, i, value[i]) != 0)
            return cc;
    }
}

if ((node_COMPLETE_DST_manual_Routes = h_class_new_node(cls_COMPLETE_DST_manual, h_category_decision, h_kind_discrete)) == 0 ||
    h_node_set_name(node_COMPLETE_DST_manual_Routes, (h_string_t)s[13]) != 0 ||
    h_node_set_number_of_states(node_COMPLETE_DST_manual_Routes, 20) != 0)
    return cc;
{
    static size_t label[] = {42, 43, 44, 45, 46, 47, 48, 49, 50, 51, 52, 53, 54, 55, 56, 57, 58, 59, 60, 61};
    for (i = 0; i < 20; ++i) {
        if (h_node_set_state_label(node_COMPLETE_DST_manual_Routes, i, (h_string_t)s[label[i]]) != 0)
            return cc;
    }
}

if ((node_COMPLETE_DST_manual_C10 = h_class_new_node(cls_COMPLETE_DST_manual, h_category_chance, h_kind_discrete)) == 0 ||
    h_node_set_name(node_COMPLETE_DST_manual_C10, (h_string_t)s[14]) != 0 ||
    h_node_set_number_of_states(node_COMPLETE_DST_manual_C10, 15) != 0 ||
    h_node_set_subtype(node_COMPLETE_DST_manual_C10, h_subtype_interval) != 0)
    return cc;
{
```

```
    static h_number_t value[] = {0.0, 2.0, 4.0, 6.0, 8.0, 10.0, 12.0, 14.0, 16.0, 18.0, 20.0, 22.0, 24.0, 26.0, 28.0, 30.0};
    for (i = 0; i <= 15; ++i) {
        if (h_node_set_state_value(node_COMPLETE_DST_manual_C10, i, value[i]) != 0)
            return cc;
    }
}

if ((node_COMPLETE_DST_manual_C6 = h_class_new_node(cls_COMPLETE_DST_manual, h_category_decision, h_kind_discrete)) == 0 ||
    h_node_set_name(node_COMPLETE_DST_manual_C6, (h_string_t)s[15]) != 0 ||
    h_node_set_number_of_states(node_COMPLETE_DST_manual_C6, 3) != 0)
    return cc;
{
    static size_t label[] = {62, 63, 64};
    for (i = 0; i < 3; ++i) {
        if (h_node_set_state_label(node_COMPLETE_DST_manual_C6, i, (h_string_t)s[label[i]]) != 0)
            return cc;
    }
}

if ((node_COMPLETE_DST_manual_D1 = h_class_new_node(cls_COMPLETE_DST_manual, h_category_decision, h_kind_discrete)) == 0 ||
    h_node_set_name(node_COMPLETE_DST_manual_D1, (h_string_t)s[16]) != 0 ||
    h_node_set_number_of_states(node_COMPLETE_DST_manual_D1, 2) != 0)
    return cc;
{
    static size_t label[] = {65, 66};
    for (i = 0; i < 2; ++i) {
        if (h_node_set_state_label(node_COMPLETE_DST_manual_D1, i, (h_string_t)s[label[i]]) != 0)
            return cc;
    }
}

if ((node_COMPLETE_DST_manual_C9 = h_class_new_node(cls_COMPLETE_DST_manual, h_category_chance, h_kind_discrete)) == 0 ||
    h_node_set_name(node_COMPLETE_DST_manual_C9, (h_string_t)s[17]) != 0 ||
    h_node_set_number_of_states(node_COMPLETE_DST_manual_C9, 2) != 0 ||
    h_node_set_subtype(node_COMPLETE_DST_manual_C9, h_subtype_number) != 0)
    return cc;
{
    static h_number_t value[] = {0.0, 7.0};
    for (i = 0; i < 2; ++i) {
        if (h_node_set_state_value(node_COMPLETE_DST_manual_C9, i, value[i]) != 0)
            return cc;
    }
}

if ((node_COMPLETE_DST_manual_U6 = h_class_new_node(cls_COMPLETE_DST_manual, h_category_utility, h_kind_other)) == 0 ||
    h_node_set_name(node_COMPLETE_DST_manual_U6, (h_string_t)s[18]) != 0)
    return cc;

if ((node_COMPLETE_DST_manual_U5 = h_class_new_node(cls_COMPLETE_DST_manual, h_category_utility, h_kind_other)) == 0 ||
    h_node_set_name(node_COMPLETE_DST_manual_U5, (h_string_t)s[19]) != 0)
    return cc;

if ((node_COMPLETE_DST_manual_C3 = h_class_new_node(cls_COMPLETE_DST_manual, h_category_chance, h_kind_discrete)) == 0 ||
    h_node_set_name(node_COMPLETE_DST_manual_C3, (h_string_t)s[20]) != 0 ||
    h_node_set_number_of_states(node_COMPLETE_DST_manual_C3, 14) != 0 ||
    h_node_set_subtype(node_COMPLETE_DST_manual_C3, h_subtype_number) != 0)
    return cc;
{
    static h_number_t value[] = {1.0, 3.0, 7.0, 9.0, 10.0, 15.0, 17.0, 18.0, 30.0, 31.0, 32.0, 33.0, 36.0, 53.0};
    for (i = 0; i < 14; ++i) {
        if (h_node_set_state_value(node_COMPLETE_DST_manual_C3, i, value[i]) != 0)
            return cc;
    }
}

if ((node_COMPLETE_DST_manual_C8 = h_class_new_node(cls_COMPLETE_DST_manual, h_category_chance, h_kind_discrete)) == 0 ||
    h_node_set_name(node_COMPLETE_DST_manual_C8, (h_string_t)s[21]) != 0 ||
    h_node_set_number_of_states(node_COMPLETE_DST_manual_C8, 12) != 0 ||
    h_node_set_subtype(node_COMPLETE_DST_manual_C8, h_subtype_interval) != 0)
    return cc;
{
    static h_number_t value[] = {0.0, 5.0E-4, 0.001, 0.005, 0.01, 0.02, 0.04, 0.06, 0.08, 0.1, 0.2, 1.0, 1.04};
    for (i = 0; i <= 12; ++i) {
        if (h_node_set_state_value(node_COMPLETE_DST_manual_C8, i, value[i]) != 0)
            return cc;
    }
}

if ((node_COMPLETE_DST_manual_C4 = h_class_new_node(cls_COMPLETE_DST_manual, h_category_chance, h_kind_discrete)) == 0 ||
    h_node_set_name(node_COMPLETE_DST_manual_C4, (h_string_t)s[22]) != 0 ||
    h_node_set_number_of_states(node_COMPLETE_DST_manual_C4, 2) != 0)
    return cc;
```

```
{
    static size_t label[] = {67, 68};
    for (i = 0; i < 2; ++i) {
        if (h_node_set_state_label(node_COMPLETE_DST_manual_C4, i, (h_string_t)s[label[i]]) != 0)
            return cc;
    }
}

if ((node_COMPLETE_DST_manual_D7 = h_class_new_node(cls_COMPLETE_DST_manual, h_category_decision, h_kind_discrete)) == 0 ||
    h_node_set_name(node_COMPLETE_DST_manual_D7, (h_string_t)s[23]) != 0 ||
    h_node_set_number_of_states(node_COMPLETE_DST_manual_D7, 4) != 0 ||
    h_node_set_subtype(node_COMPLETE_DST_manual_D7, h_subtype_interval) != 0)
    return cc;
{
    static h_number_t value[] = {1000.0, 2000.0, 3000.0, 4000.0, 5000.0};
    for (i = 0; i <= 4; ++i) {
        if (h_node_set_state_value(node_COMPLETE_DST_manual_D7, i, value[i]) != 0)
            return cc;
    }
}

if ((node_COMPLETE_DST_manual_D6 = h_class_new_node(cls_COMPLETE_DST_manual, h_category_decision, h_kind_discrete)) == 0 ||
    h_node_set_name(node_COMPLETE_DST_manual_D6, (h_string_t)s[24]) != 0 ||
    h_node_set_number_of_states(node_COMPLETE_DST_manual_D6, 2) != 0)
    return cc;
{
    static size_t label[] = {69, 70};
    for (i = 0; i < 2; ++i) {
        if (h_node_set_state_label(node_COMPLETE_DST_manual_D6, i, (h_string_t)s[label[i]]) != 0)
            return cc;
    }
}

if ((node_COMPLETE_DST_manual_C12 = h_class_new_node(cls_COMPLETE_DST_manual, h_category_decision, h_kind_discrete)) == 0 ||
    h_node_set_name(node_COMPLETE_DST_manual_C12, (h_string_t)s[25]) != 0 ||
    h_node_set_number_of_states(node_COMPLETE_DST_manual_C12, 5) != 0 ||
    h_node_set_subtype(node_COMPLETE_DST_manual_C12, h_subtype_number) != 0)
    return cc;
{
    static h_number_t value[] = {0.0, 1.0, 2.0, 3.0, 4.0};
    for (i = 0; i < 5; ++i) {
        if (h_node_set_state_value(node_COMPLETE_DST_manual_C12, i, value[i]) != 0)
            return cc;
    }
}

if ((node_COMPLETE_DST_manual_U4 = h_class_new_node(cls_COMPLETE_DST_manual, h_category_utility, h_kind_other)) == 0 ||
    h_node_set_name(node_COMPLETE_DST_manual_U4, (h_string_t)s[26]) != 0)
    return cc;

if ((node_COMPLETE_DST_manual_C20 = h_class_new_node(cls_COMPLETE_DST_manual, h_category_chance, h_kind_discrete)) == 0 ||
    h_node_set_name(node_COMPLETE_DST_manual_C20, (h_string_t)s[27]) != 0 ||
    h_node_set_number_of_states(node_COMPLETE_DST_manual_C20, 13) != 0 ||
    h_node_set_subtype(node_COMPLETE_DST_manual_C20, h_subtype_interval) != 0)
    return cc;
{
    static h_number_t value[] = {2000.0, 3000.0, 4000.0, 5000.0, 6000.0, 8000.0, 10000.0, 12000.0, 14000.0, 16000.0, 18000.0, 20000.0, 22000.0, 24000.0};
    for (i = 0; i <= 13; ++i) {
        if (h_node_set_state_value(node_COMPLETE_DST_manual_C20, i, value[i]) != 0)
            return cc;
    }
}

if ((node_COMPLETE_DST_manual_D3 = h_class_new_node(cls_COMPLETE_DST_manual, h_category_decision, h_kind_discrete)) == 0 ||
    h_node_set_name(node_COMPLETE_DST_manual_D3, (h_string_t)s[28]) != 0 ||
    h_node_set_number_of_states(node_COMPLETE_DST_manual_D3, 6) != 0)
    return cc;
{
    static size_t label[] = {71, 72, 73, 74, 75, 76};
    for (i = 0; i < 6; ++i) {
        if (h_node_set_state_label(node_COMPLETE_DST_manual_D3, i, (h_string_t)s[label[i]]) != 0)
            return cc;
    }
}

if ((node_COMPLETE_DST_manual_C11 = h_class_new_node(cls_COMPLETE_DST_manual, h_category_chance, h_kind_discrete)) == 0 ||
    h_node_set_name(node_COMPLETE_DST_manual_C11, (h_string_t)s[29]) != 0 ||
    h_node_set_number_of_states(node_COMPLETE_DST_manual_C11, 6) != 0 ||
    h_node_set_subtype(node_COMPLETE_DST_manual_C11, h_subtype_interval) != 0)
    return cc;
{
    static h_number_t value[] = {1000.0, 2000.0, 3000.0, 4000.0, 5000.0, 6000.0, 7000.0};
```

```
    for (i = 0; i <= 6; ++i) {
        if (h_node_set_state_value(node_COMPLETE_DST_manual_C11, i, value[i]) != 0)
            return cc;
    }
}

if ((node_COMPLETE_DST_manual_U3 = h_class_new_node(cls_COMPLETE_DST_manual, h_category_utility, h_kind_other)) == 0 ||
    h_node_set_name(node_COMPLETE_DST_manual_U3, (h_string_t)s[30]) != 0)
    return cc;

if ((node_COMPLETE_DST_manual_U2 = h_class_new_node(cls_COMPLETE_DST_manual, h_category_utility, h_kind_other)) == 0 ||
    h_node_set_name(node_COMPLETE_DST_manual_U2, (h_string_t)s[31]) != 0)
    return cc;

if ((node_COMPLETE_DST_manual_U1 = h_class_new_node(cls_COMPLETE_DST_manual, h_category_utility, h_kind_other)) == 0 ||
    h_node_set_name(node_COMPLETE_DST_manual_U1, (h_string_t)s[32]) != 0)
    return cc;

if ((node_COMPLETE_DST_manual_C5 = h_class_new_node(cls_COMPLETE_DST_manual, h_category_decision, h_kind_discrete)) == 0 ||
    h_node_set_name(node_COMPLETE_DST_manual_C5, (h_string_t)s[33]) != 0 ||
    h_node_set_number_of_states(node_COMPLETE_DST_manual_C5, 4) != 0 ||
    h_node_set_subtype(node_COMPLETE_DST_manual_C5, h_subtype_number) != 0)
    return cc;
{
    static h_number_t value[] = {0.0, 2.0, 6.0, 12.0};
    for (i = 0; i < 4; ++i) {
        if (h_node_set_state_value(node_COMPLETE_DST_manual_C5, i, value[i]) != 0)
            return cc;
    }
}

//structure class COMPLETE_DST_manual
if (h_node_add_parent(node_COMPLETE_DST_manual_U9, node_COMPLETE_DST_manual_D2) != 0)
    return cc;
if (h_node_add_parent(node_COMPLETE_DST_manual_U9, node_COMPLETE_DST_manual_C20) != 0)
    return cc;
if (h_node_add_parent(node_COMPLETE_DST_manual_C17, node_COMPLETE_DST_manual_C16) != 0)
    return cc;
if (h_node_add_parent(node_COMPLETE_DST_manual_C17, node_COMPLETE_DST_manual_C12) != 0)
    return cc;
if (h_node_add_parent(node_COMPLETE_DST_manual_C17, node_COMPLETE_DST_manual_C5) != 0)
    return cc;
if (h_node_add_parent(node_COMPLETE_DST_manual_C16, node_COMPLETE_DST_manual_Routes) != 0)
    return cc;
if (h_node_add_parent(node_COMPLETE_DST_manual_C1, node_COMPLETE_DST_manual_C16) != 0)
    return cc;
if (h_node_add_parent(node_COMPLETE_DST_manual_C1, node_COMPLETE_DST_manual_C12) != 0)
    return cc;
if (h_node_add_parent(node_COMPLETE_DST_manual_C1, node_COMPLETE_DST_manual_C5) != 0)
    return cc;
if (h_node_add_parent(node_COMPLETE_DST_manual_U8, node_COMPLETE_DST_manual_C11) != 0)
    return cc;
if (h_node_add_parent(node_COMPLETE_DST_manual_U8, node_COMPLETE_DST_manual_D6) != 0)
    return cc;
if (h_node_add_parent(node_COMPLETE_DST_manual_U8, node_COMPLETE_DST_manual_D2) != 0)
    return cc;
if (h_node_add_parent(node_COMPLETE_DST_manual_C21, node_COMPLETE_DST_manual_D1) != 0)
    return cc;
if (h_node_add_parent(node_COMPLETE_DST_manual_C21, node_COMPLETE_DST_manual_C5) != 0)
    return cc;
if (h_node_add_parent(node_COMPLETE_DST_manual_C21, node_COMPLETE_DST_manual_C9) != 0)
    return cc;
if (h_node_add_parent(node_COMPLETE_DST_manual_C21, node_COMPLETE_DST_manual_C8) != 0)
    return cc;
if (h_node_add_parent(node_COMPLETE_DST_manual_C19, node_COMPLETE_DST_manual_C8) != 0)
    return cc;
if (h_node_add_parent(node_COMPLETE_DST_manual_C19, node_COMPLETE_DST_manual_D3) != 0)
    return cc;
if (h_node_add_parent(node_COMPLETE_DST_manual_U7, node_COMPLETE_DST_manual_C4) != 0)
    return cc;
if (h_node_add_parent(node_COMPLETE_DST_manual_U7, node_COMPLETE_DST_manual_C16) != 0)
    return cc;
if (h_node_add_parent(node_COMPLETE_DST_manual_C18, node_COMPLETE_DST_manual_C17) != 0)
    return cc;
if (h_node_add_parent(node_COMPLETE_DST_manual_C14, node_COMPLETE_DST_manual_C10) != 0)
    return cc;
if (h_node_add_parent(node_COMPLETE_DST_manual_C14, node_COMPLETE_DST_manual_C3) != 0)
    return cc;
if (h_node_add_parent(node_COMPLETE_DST_manual_C14, node_COMPLETE_DST_manual_C17) != 0)
    return cc;
if (h_node_add_parent(node_COMPLETE_DST_manual_C13, node_COMPLETE_DST_manual_C3) != 0)
    return cc;
```

```
if (h_node_add_parent(node_COMPLETE_DST_manual_C13, node_COMPLETE_DST_manual_C19) != 0)
    return cc;
if (h_node_add_parent(node_COMPLETE_DST_manual_C13, node_COMPLETE_DST_manual_C17) != 0)
    return cc;
if (h_node_add_parent(node_COMPLETE_DST_manual_C10, node_COMPLETE_DST_manual_C8) != 0)
    return cc;
if (h_node_add_parent(node_COMPLETE_DST_manual_C10, node_COMPLETE_DST_manual_D3) != 0)
    return cc;
if (h_node_add_parent(node_COMPLETE_DST_manual_C9, node_COMPLETE_DST_manual_C16) != 0)
    return cc;
if (h_node_add_parent(node_COMPLETE_DST_manual_U6, node_COMPLETE_DST_manual_C8) != 0)
    return cc;
if (h_node_add_parent(node_COMPLETE_DST_manual_U6, node_COMPLETE_DST_manual_C16) != 0)
    return cc;
if (h_node_add_parent(node_COMPLETE_DST_manual_U5, node_COMPLETE_DST_manual_C5) != 0)
    return cc;
if (h_node_add_parent(node_COMPLETE_DST_manual_U5, node_COMPLETE_DST_manual_C8) != 0)
    return cc;
if (h_node_add_parent(node_COMPLETE_DST_manual_U5, node_COMPLETE_DST_manual_D4) != 0)
    return cc;
if (h_node_add_parent(node_COMPLETE_DST_manual_U5, node_COMPLETE_DST_manual_C6) != 0)
    return cc;
if (h_node_add_parent(node_COMPLETE_DST_manual_C3, node_COMPLETE_DST_manual_Routes) != 0)
    return cc;
if (h_node_add_parent(node_COMPLETE_DST_manual_C8, node_COMPLETE_DST_manual_D3) != 0)
    return cc;
if (h_node_add_parent(node_COMPLETE_DST_manual_C4, node_COMPLETE_DST_manual_Routes) != 0)
    return cc;
if (h_node_add_parent(node_COMPLETE_DST_manual_U4, node_COMPLETE_DST_manual_C20) != 0)
    return cc;
if (h_node_add_parent(node_COMPLETE_DST_manual_C20, node_COMPLETE_DST_manual_C11) != 0)
    return cc;
if (h_node_add_parent(node_COMPLETE_DST_manual_C20, node_COMPLETE_DST_manual_D6) != 0)
    return cc;
if (h_node_add_parent(node_COMPLETE_DST_manual_C11, node_COMPLETE_DST_manual_D7) != 0)
    return cc;
if (h_node_add_parent(node_COMPLETE_DST_manual_C11, node_COMPLETE_DST_manual_C1) != 0)
    return cc;
if (h_node_add_parent(node_COMPLETE_DST_manual_U3, node_COMPLETE_DST_manual_C13) != 0)
    return cc;
if (h_node_add_parent(node_COMPLETE_DST_manual_U3, node_COMPLETE_DST_manual_C14) != 0)
    return cc;
if (h_node_add_parent(node_COMPLETE_DST_manual_U3, node_COMPLETE_DST_manual_C6) != 0)
    return cc;
if (h_node_add_parent(node_COMPLETE_DST_manual_U3, node_COMPLETE_DST_manual_C18) != 0)
    return cc;
if (h_node_add_parent(node_COMPLETE_DST_manual_U2, node_COMPLETE_DST_manual_C21) != 0)
    return cc;
if (h_node_add_parent(node_COMPLETE_DST_manual_U1, node_COMPLETE_DST_manual_C11) != 0)
    return cc;
if (h_node_add_parent(node_COMPLETE_DST_manual_U1, node_COMPLETE_DST_manual_D6) != 0)
    return cc;

//parameters class COMPLETE_DST_manual

node_buffer[0] = 0;
if ((model = h_node_new_model(node_COMPLETE_DST_manual_U9, node_buffer)) == 0)
    return cc;
if ((expr = h_string_parse_expression((h_string_t)s[77], model, 0, 0)) == 0)
    return cc;
if (h_model_set_expression(model, 0, expr) != 0)
    return cc;

if ((t = h_node_get_table(node_COMPLETE_DST_manual_C17)) == 0 || (d = h_table_get_data(t)) == 0)
    return cc;
{
    static h_number_t data[] = {1.0,0.0,1.0,0.7,0.3,0.6,0.4,0.4,0.6};
    memcpy(d+1, arr_0_0, sizeof(h_number_t)*5);
    memcpy(d+7, arr_0_0, sizeof(h_number_t)*5);
    memcpy(d+13, arr_0_0, sizeof(h_number_t)*10);
    memcpy(d+25, arr_0_0, sizeof(h_number_t)*5);
    memcpy(d+31, arr_0_0, sizeof(h_number_t)*10);
    memcpy(d+42, arr_0_0, sizeof(h_number_t)*2);
    memcpy(d+45, arr_0_0, sizeof(h_number_t)*3);
    memcpy(d+49, arr_0_0, sizeof(h_number_t)*10);
    memcpy(d+60, arr_0_0, sizeof(h_number_t)*5);
    memcpy(d+67, arr_0_0, sizeof(h_number_t)*10);
    memcpy(d+78, arr_0_0, sizeof(h_number_t)*5);
    memcpy(d+85, arr_0_0, sizeof(h_number_t)*5);
    memcpy(d+91, arr_0_0, sizeof(h_number_t)*5);
    memcpy(d+97, arr_0_0, sizeof(h_number_t)*5);
    memcpy(d+103, arr_0_0, sizeof(h_number_t)*10);
```

```
    memcpy(d+115, arr_0_0, sizeof(h_number_t)*5);
    memcpy(d+121, arr_0_0, sizeof(h_number_t)*10);
    memcpy(d+131, data, sizeof(h_number_t)*3);
    memcpy(d+134, arr_0_0, sizeof(h_number_t)*4);
    memcpy(d+139, arr_0_0, sizeof(h_number_t)*10);
    memcpy(d+150, arr_0_0, sizeof(h_number_t)*2);
    memcpy(d+153, arr_0_0, sizeof(h_number_t)*3);
    memcpy(d+157, arr_0_0, sizeof(h_number_t)*10);
    memcpy(d+168, arr_0_0, sizeof(h_number_t)*2);
    memcpy(d+171, arr_0_0, sizeof(h_number_t)*3);
    memcpy(d+175, arr_0_0, sizeof(h_number_t)*5);
    memcpy(d+181, arr_0_0, sizeof(h_number_t)*5);
    memcpy(d+187, arr_0_0, sizeof(h_number_t)*5);
    memcpy(d+193, arr_0_0, sizeof(h_number_t)*6);
    memcpy(d+199, data+3, sizeof(h_number_t)*2);
    memcpy(d+201, arr_0_0, sizeof(h_number_t)*3);
    memcpy(d+205, arr_0_0, sizeof(h_number_t)*5);
    memcpy(d+211, arr_0_0, sizeof(h_number_t)*6);
    memcpy(d+217, data+5, sizeof(h_number_t)*2);
    memcpy(d+219, arr_0_0, sizeof(h_number_t)*3);
    memcpy(d+223, arr_0_0, sizeof(h_number_t)*5);
    memcpy(d+229, arr_0_0, sizeof(h_number_t)*6);
    memcpy(d+235, arr_0_5, sizeof(h_number_t)*2);
    memcpy(d+237, arr_0_0, sizeof(h_number_t)*3);
    memcpy(d+241, arr_0_0, sizeof(h_number_t)*5);
    memcpy(d+247, arr_0_0, sizeof(h_number_t)*6);
    memcpy(d+253, data+7, sizeof(h_number_t)*2);
    memcpy(d+255, arr_0_0, sizeof(h_number_t)*3);
    memcpy(d+259, arr_0_0, sizeof(h_number_t)*5);
    memcpy(d+265, arr_0_0, sizeof(h_number_t)*5);
    memcpy(d+271, arr_0_0, sizeof(h_number_t)*5);
    memcpy(d+277, arr_0_0, sizeof(h_number_t)*5);
    memcpy(d+283, arr_0_0, sizeof(h_number_t)*5);
    memcpy(d+289, arr_0_0, sizeof(h_number_t)*5);
    memcpy(d+295, arr_0_0, sizeof(h_number_t)*5);
    memcpy(d+301, arr_0_0, sizeof(h_number_t)*5);
    memcpy(d+307, arr_0_0, sizeof(h_number_t)*5);
    memcpy(d+313, arr_0_0, sizeof(h_number_t)*5);
    memcpy(d+319, arr_0_0, sizeof(h_number_t)*5);
    memcpy(d+325, arr_0_0, sizeof(h_number_t)*5);
    memcpy(d+331, arr_0_0, sizeof(h_number_t)*5);
    memcpy(d+337, arr_0_0, sizeof(h_number_t)*5);
    memcpy(d+343, arr_0_0, sizeof(h_number_t)*5);
    memcpy(d+349, arr_0_0, sizeof(h_number_t)*5);
    memcpy(d+355, arr_0_0, sizeof(h_number_t)*5);
}

if ((t = h_node_get_table(node_COMPLETE_DST_manual_C16)) == 0 || (d = h_table_get_data(t)) == 0)
    return cc;
{
    static h_number_t data[] = {0.0,0.0,0.0,0.0,0.0,0.0,0.0,0.0,0.0,0.0,0.0,0.0,0.0,0.0,0.0,0.0,0.0,0.0};
    memcpy(d, arr_0_3, sizeof(h_number_t)*6);
    memcpy(d+6, arr_0_5, sizeof(h_number_t)*2);
    memcpy(d+8, data, sizeof(h_number_t));
    memcpy(d+9, arr_0_5, sizeof(h_number_t)*2);
    memcpy(d+11, data+1, sizeof(h_number_t));
    memcpy(d+12, arr_0_5, sizeof(h_number_t)*2);
    memcpy(d+14, data+2, sizeof(h_number_t));
    memcpy(d+15, arr_0_5, sizeof(h_number_t)*2);
    memcpy(d+17, data+3, sizeof(h_number_t));
    memcpy(d+18, arr_0_5, sizeof(h_number_t)*2);
    memcpy(d+20, data+4, sizeof(h_number_t));
    memcpy(d+21, arr_0_5, sizeof(h_number_t)*2);
    memcpy(d+23, data+5, sizeof(h_number_t));
    memcpy(d+24, arr_0_5, sizeof(h_number_t)*2);
    memcpy(d+26, data+6, sizeof(h_number_t));
    memcpy(d+27, arr_0_5, sizeof(h_number_t)*2);
    memcpy(d+29, data+7, sizeof(h_number_t));
    memcpy(d+30, arr_0_5, sizeof(h_number_t)*2);
    memcpy(d+32, data+8, sizeof(h_number_t));
    memcpy(d+33, arr_0_5, sizeof(h_number_t)*2);
    memcpy(d+35, data+9, sizeof(h_number_t));
    memcpy(d+36, arr_0_5, sizeof(h_number_t)*2);
    memcpy(d+38, data+10, sizeof(h_number_t));
    memcpy(d+39, arr_0_5, sizeof(h_number_t)*2);
    memcpy(d+41, data+11, sizeof(h_number_t));
    memcpy(d+42, arr_0_5, sizeof(h_number_t)*2);
    memcpy(d+44, data+12, sizeof(h_number_t));
    memcpy(d+45, arr_0_5, sizeof(h_number_t)*2);
    memcpy(d+47, data+13, sizeof(h_number_t));
    memcpy(d+48, arr_0_5, sizeof(h_number_t)*2);
    memcpy(d+50, data+14, sizeof(h_number_t));
```

```
    memcpy(d+51, arr_0_5, sizeof(h_number_t)*2);
    memcpy(d+53, data+15, sizeof(h_number_t));
    memcpy(d+54, arr_0_3, sizeof(h_number_t)*6);
}

if ((t = h_node_get_table(node_COMPLETE_DST_manual_C1)) == 0 || (d = h_table_get_data(t)) == 0)
    return cc;
{
    static h_number_t data[] = {1.0,0.0,1.0,1.0,0.0,1.0,0.0,1.0,1.0,0.0,1.0,1.0,0.0,1.0,0.7,0.3,0.6,0.4,0.4,0.6};
    memcpy(d+1, arr_0_0, sizeof(h_number_t)*5);
    memcpy(d+7, arr_0_0, sizeof(h_number_t)*5);
    memcpy(d+13, arr_0_0, sizeof(h_number_t)*10);
    memcpy(d+25, arr_0_0, sizeof(h_number_t)*5);
    memcpy(d+31, arr_0_0, sizeof(h_number_t)*10);
    memcpy(d+41, data, sizeof(h_number_t)*3);
    memcpy(d+44, arr_0_0, sizeof(h_number_t)*4);
    memcpy(d+49, arr_0_0, sizeof(h_number_t)*10);
    memcpy(d+60, arr_0_0, sizeof(h_number_t)*4);
    memcpy(d+64, data+3, sizeof(h_number_t)*3);
    memcpy(d+67, arr_0_0, sizeof(h_number_t)*10);
    memcpy(d+78, arr_0_0, sizeof(h_number_t)*5);
    memcpy(d+85, arr_0_0, sizeof(h_number_t)*5);
    memcpy(d+91, arr_0_0, sizeof(h_number_t)*5);
    memcpy(d+97, arr_0_0, sizeof(h_number_t)*5);
    memcpy(d+103, arr_0_0, sizeof(h_number_t)*8);
    memcpy(d+111, arr_0_5, sizeof(h_number_t)*2);
    memcpy(d+113, data+6, sizeof(h_number_t)*2);
    memcpy(d+115, arr_0_0, sizeof(h_number_t)*5);
    memcpy(d+121, arr_0_0, sizeof(h_number_t)*10);
    memcpy(d+133, arr_0_0, sizeof(h_number_t)*5);
    memcpy(d+139, arr_0_0, sizeof(h_number_t)*10);
    memcpy(d+149, data+8, sizeof(h_number_t)*3);
    memcpy(d+152, arr_0_0, sizeof(h_number_t)*4);
    memcpy(d+157, arr_0_0, sizeof(h_number_t)*10);
    memcpy(d+167, data+11, sizeof(h_number_t)*3);
    memcpy(d+170, arr_0_0, sizeof(h_number_t)*4);
    memcpy(d+175, arr_0_0, sizeof(h_number_t)*5);
    memcpy(d+181, arr_0_0, sizeof(h_number_t)*5);
    memcpy(d+187, arr_0_0, sizeof(h_number_t)*5);
    memcpy(d+193, arr_0_0, sizeof(h_number_t)*6);
    memcpy(d+199, data+14, sizeof(h_number_t)*2);
    memcpy(d+201, arr_0_0, sizeof(h_number_t)*3);
    memcpy(d+205, arr_0_0, sizeof(h_number_t)*5);
    memcpy(d+211, arr_0_0, sizeof(h_number_t)*6);
    memcpy(d+217, data+16, sizeof(h_number_t)*2);
    memcpy(d+219, arr_0_0, sizeof(h_number_t)*3);
    memcpy(d+223, arr_0_0, sizeof(h_number_t)*5);
    memcpy(d+229, arr_0_0, sizeof(h_number_t)*6);
    memcpy(d+235, arr_0_5, sizeof(h_number_t)*2);
    memcpy(d+237, arr_0_0, sizeof(h_number_t)*3);
    memcpy(d+241, arr_0_0, sizeof(h_number_t)*5);
    memcpy(d+247, arr_0_0, sizeof(h_number_t)*6);
    memcpy(d+253, data+18, sizeof(h_number_t)*2);
    memcpy(d+255, arr_0_0, sizeof(h_number_t)*3);
    memcpy(d+259, arr_0_0, sizeof(h_number_t)*5);
    memcpy(d+265, arr_0_0, sizeof(h_number_t)*5);
    memcpy(d+271, arr_0_0, sizeof(h_number_t)*5);
    memcpy(d+277, arr_0_0, sizeof(h_number_t)*5);
    memcpy(d+283, arr_0_0, sizeof(h_number_t)*5);
    memcpy(d+289, arr_0_0, sizeof(h_number_t)*5);
    memcpy(d+295, arr_0_0, sizeof(h_number_t)*5);
    memcpy(d+301, arr_0_0, sizeof(h_number_t)*5);
    memcpy(d+307, arr_0_0, sizeof(h_number_t)*5);
    memcpy(d+313, arr_0_0, sizeof(h_number_t)*5);
    memcpy(d+319, arr_0_0, sizeof(h_number_t)*5);
    memcpy(d+325, arr_0_0, sizeof(h_number_t)*5);
    memcpy(d+331, arr_0_0, sizeof(h_number_t)*5);
    memcpy(d+337, arr_0_0, sizeof(h_number_t)*5);
    memcpy(d+343, arr_0_0, sizeof(h_number_t)*5);
    memcpy(d+349, arr_0_0, sizeof(h_number_t)*5);
    memcpy(d+355, arr_0_0, sizeof(h_number_t)*5);
}

node_buffer[0] = node_COMPLETE_DST_manual_D6;
node_buffer[1] = 0;
if ((model = h_node_new_model(node_COMPLETE_DST_manual_U8, node_buffer)) == 0)
    return cc;
if ((expr = h_string_parse_expression((h_string_t)s[78], model, 0, 0)) == 0)
    return cc;
if (h_model_set_expression(model, 0, expr) != 0)
    return cc;
if ((expr = h_string_parse_expression((h_string_t)s[79], model, 0, 0)) == 0)
```

```
    return cc;
if (h_model_set_expression(model, 1, expr) != 0)
    return cc;

node_buffer[0] = node_COMPLETE_DST_manual_D1;
node_buffer[1] = node_COMPLETE_DST_manual_C9;
node_buffer[2] = node_COMPLETE_DST_manual_C5;
node_buffer[3] = 0;
if ((model = h_node_new_model(node_COMPLETE_DST_manual_C21, node_buffer)) == 0)
    return cc;
if ((expr = h_string_parse_expression((h_string_t)s[80], model, 0, 0)) == 0)
    return cc;
if (h_model_set_expression(model, 0, expr) != 0)
    return cc;
if ((expr = h_string_parse_expression((h_string_t)s[80], model, 0, 0)) == 0)
    return cc;
if (h_model_set_expression(model, 1, expr) != 0)
    return cc;
if ((expr = h_string_parse_expression((h_string_t)s[80], model, 0, 0)) == 0)
    return cc;
if (h_model_set_expression(model, 2, expr) != 0)
    return cc;
if ((expr = h_string_parse_expression((h_string_t)s[80], model, 0, 0)) == 0)
    return cc;
if (h_model_set_expression(model, 3, expr) != 0)
    return cc;
if ((expr = h_string_parse_expression((h_string_t)s[81], model, 0, 0)) == 0)
    return cc;
if (h_model_set_expression(model, 4, expr) != 0)
    return cc;
if ((expr = h_string_parse_expression((h_string_t)s[82], model, 0, 0)) == 0)
    return cc;
if (h_model_set_expression(model, 5, expr) != 0)
    return cc;
if ((expr = h_string_parse_expression((h_string_t)s[83], model, 0, 0)) == 0)
    return cc;
if (h_model_set_expression(model, 6, expr) != 0)
    return cc;
if ((expr = h_string_parse_expression((h_string_t)s[84], model, 0, 0)) == 0)
    return cc;
if (h_model_set_expression(model, 7, expr) != 0)
    return cc;
if ((expr = h_string_parse_expression((h_string_t)s[80], model, 0, 0)) == 0)
    return cc;
if (h_model_set_expression(model, 8, expr) != 0)
    return cc;
if ((expr = h_string_parse_expression((h_string_t)s[80], model, 0, 0)) == 0)
    return cc;
if (h_model_set_expression(model, 9, expr) != 0)
    return cc;
if ((expr = h_string_parse_expression((h_string_t)s[80], model, 0, 0)) == 0)
    return cc;
if (h_model_set_expression(model, 10, expr) != 0)
    return cc;
if ((expr = h_string_parse_expression((h_string_t)s[80], model, 0, 0)) == 0)
    return cc;
if (h_model_set_expression(model, 11, expr) != 0)
    return cc;
if ((expr = h_string_parse_expression((h_string_t)s[81], model, 0, 0)) == 0)
    return cc;
if (h_model_set_expression(model, 12, expr) != 0)
    return cc;
if ((expr = h_string_parse_expression((h_string_t)s[85], model, 0, 0)) == 0)
    return cc;
if (h_model_set_expression(model, 13, expr) != 0)
    return cc;
if ((expr = h_string_parse_expression((h_string_t)s[86], model, 0, 0)) == 0)
    return cc;
if (h_model_set_expression(model, 14, expr) != 0)
    return cc;
if ((expr = h_string_parse_expression((h_string_t)s[87], model, 0, 0)) == 0)
    return cc;
if (h_model_set_expression(model, 15, expr) != 0)
    return cc;

node_buffer[0] = node_COMPLETE_DST_manual_D3;
node_buffer[1] = 0;
if ((model = h_node_new_model(node_COMPLETE_DST_manual_C19, node_buffer)) == 0)
    return cc;
if ((expr = h_string_parse_expression((h_string_t)s[88], model, 0, 0)) == 0)
    return cc;
if (h_model_set_expression(model, 0, expr) != 0)
```

```
    return cc;
if ((expr = h_string_parse_expression((h_string_t)s[89], model, 0, 0)) == 0)
    return cc;
if (h_model_set_expression(model, 1, expr) != 0)
    return cc;
if ((expr = h_string_parse_expression((h_string_t)s[89], model, 0, 0)) == 0)
    return cc;
if (h_model_set_expression(model, 2, expr) != 0)
    return cc;
if ((expr = h_string_parse_expression((h_string_t)s[90], model, 0, 0)) == 0)
    return cc;
if (h_model_set_expression(model, 3, expr) != 0)
    return cc;
if ((expr = h_string_parse_expression((h_string_t)s[90], model, 0, 0)) == 0)
    return cc;
if (h_model_set_expression(model, 4, expr) != 0)
    return cc;
if ((expr = h_string_parse_expression((h_string_t)s[91], model, 0, 0)) == 0)
    return cc;
if (h_model_set_expression(model, 5, expr) != 0)
    return cc;

if ((t = h_node_get_table(node_COMPLETE_DST_manual_U7)) == 0 || (d = h_table_get_data(t)) == 0)
    return cc;
{
    static h_number_t data[] = {-50.0,-100.0};
    memcpy(d+2, data, sizeof(h_number_t)*2);
}

if ((t = h_node_get_table(node_COMPLETE_DST_manual_C18)) == 0 || (d = h_table_get_data(t)) == 0)
    return cc;
{
    static h_number_t data[] = {1.0,0.0,1.0,0.0,1.0,0.0,0.0,1.0,0.0,1.0};
    memcpy(d, data, sizeof(h_number_t)*6);
    memcpy(d+6, arr_0_5, sizeof(h_number_t)*2);
    memcpy(d+8, data+6, sizeof(h_number_t)*4);
}

node_buffer[0] = 0;
if ((model = h_node_new_model(node_COMPLETE_DST_manual_C14, node_buffer)) == 0)
    return cc;
if ((expr = h_string_parse_expression((h_string_t)s[92], model, 0, 0)) == 0)
    return cc;
if (h_model_set_expression(model, 0, expr) != 0)
    return cc;

if ((model = h_node_new_model(node_COMPLETE_DST_manual_C13, node_buffer)) == 0)
    return cc;
if ((expr = h_string_parse_expression((h_string_t)s[93], model, 0, 0)) == 0)
    return cc;
if (h_model_set_expression(model, 0, expr) != 0)
    return cc;

node_buffer[0] = node_COMPLETE_DST_manual_D3;
node_buffer[1] = 0;
if ((model = h_node_new_model(node_COMPLETE_DST_manual_C10, node_buffer)) == 0)
    return cc;
if ((expr = h_string_parse_expression((h_string_t)s[94], model, 0, 0)) == 0)
    return cc;
if (h_model_set_expression(model, 0, expr) != 0)
    return cc;
if ((expr = h_string_parse_expression((h_string_t)s[95], model, 0, 0)) == 0)
    return cc;
if (h_model_set_expression(model, 1, expr) != 0)
    return cc;
if ((expr = h_string_parse_expression((h_string_t)s[95], model, 0, 0)) == 0)
    return cc;
if (h_model_set_expression(model, 2, expr) != 0)
    return cc;
if ((expr = h_string_parse_expression((h_string_t)s[96], model, 0, 0)) == 0)
    return cc;
if (h_model_set_expression(model, 3, expr) != 0)
    return cc;
if ((expr = h_string_parse_expression((h_string_t)s[97], model, 0, 0)) == 0)
    return cc;
if (h_model_set_expression(model, 4, expr) != 0)
    return cc;
if ((expr = h_string_parse_expression((h_string_t)s[98], model, 0, 0)) == 0)
    return cc;
if (h_model_set_expression(model, 5, expr) != 0)
    return cc;
```

```
if ((t = h_node_get_table(node_COMPLETE_DST_manual_C9)) == 0 || (d = h_table_get_data(t)) == 0)
    return cc;
{
    static h_number_t data[] = {0.0};
    memcpy(d+1, arr_0_0, sizeof(h_number_t)*2);
    memcpy(d+5, data, sizeof(h_number_t));
}

node_buffer[0] = node_COMPLETE_DST_manual_C16;
node_buffer[1] = 0;
if ((model = h_node_new_model(node_COMPLETE_DST_manual_U6, node_buffer)) == 0)
    return cc;
if ((expr = h_string_parse_expression((h_string_t)s[99], model, 0, 0)) == 0)
    return cc;
if (h_model_set_expression(model, 0, expr) != 0)
    return cc;
if ((expr = h_string_parse_expression((h_string_t)s[100], model, 0, 0)) == 0)
    return cc;
if (h_model_set_expression(model, 1, expr) != 0)
    return cc;
if ((expr = h_string_parse_expression((h_string_t)s[101], model, 0, 0)) == 0)
    return cc;
if (h_model_set_expression(model, 2, expr) != 0)
    return cc;

node_buffer[0] = node_COMPLETE_DST_manual_C6;
node_buffer[1] = node_COMPLETE_DST_manual_D4;
node_buffer[2] = node_COMPLETE_DST_manual_C5;
node_buffer[3] = 0;
if ((model = h_node_new_model(node_COMPLETE_DST_manual_U5, node_buffer)) == 0)
    return cc;
if ((expr = h_string_parse_expression((h_string_t)s[80], model, 0, 0)) == 0)
    return cc;
if (h_model_set_expression(model, 0, expr) != 0)
    return cc;
if ((expr = h_string_parse_expression((h_string_t)s[102], model, 0, 0)) == 0)
    return cc;
if (h_model_set_expression(model, 1, expr) != 0)
    return cc;
if ((expr = h_string_parse_expression((h_string_t)s[103], model, 0, 0)) == 0)
    return cc;
if (h_model_set_expression(model, 2, expr) != 0)
    return cc;
if ((expr = h_string_parse_expression((h_string_t)s[104], model, 0, 0)) == 0)
    return cc;
if (h_model_set_expression(model, 3, expr) != 0)
    return cc;
if ((expr = h_string_parse_expression((h_string_t)s[80], model, 0, 0)) == 0)
    return cc;
if (h_model_set_expression(model, 4, expr) != 0)
    return cc;
if ((expr = h_string_parse_expression((h_string_t)s[105], model, 0, 0)) == 0)
    return cc;
if (h_model_set_expression(model, 5, expr) != 0)
    return cc;
if ((expr = h_string_parse_expression((h_string_t)s[106], model, 0, 0)) == 0)
    return cc;
if (h_model_set_expression(model, 6, expr) != 0)
    return cc;
if ((expr = h_string_parse_expression((h_string_t)s[107], model, 0, 0)) == 0)
    return cc;
if (h_model_set_expression(model, 7, expr) != 0)
    return cc;
if ((expr = h_string_parse_expression((h_string_t)s[80], model, 0, 0)) == 0)
    return cc;
if (h_model_set_expression(model, 8, expr) != 0)
    return cc;
if ((expr = h_string_parse_expression((h_string_t)s[108], model, 0, 0)) == 0)
    return cc;
if (h_model_set_expression(model, 9, expr) != 0)
    return cc;
if ((expr = h_string_parse_expression((h_string_t)s[109], model, 0, 0)) == 0)
    return cc;
if (h_model_set_expression(model, 10, expr) != 0)
    return cc;
if ((expr = h_string_parse_expression((h_string_t)s[110], model, 0, 0)) == 0)
    return cc;
if (h_model_set_expression(model, 11, expr) != 0)
    return cc;
if ((expr = h_string_parse_expression((h_string_t)s[80], model, 0, 0)) == 0)
    return cc;
if (h_model_set_expression(model, 12, expr) != 0)
```

```
    return cc;
if ((expr = h_string_parse_expression((h_string_t)s[111], model, 0, 0)) == 0)
    return cc;
if (h_model_set_expression(model, 13, expr) != 0)
    return cc;
if ((expr = h_string_parse_expression((h_string_t)s[112], model, 0, 0)) == 0)
    return cc;
if (h_model_set_expression(model, 14, expr) != 0)
    return cc;
if ((expr = h_string_parse_expression((h_string_t)s[113], model, 0, 0)) == 0)
    return cc;
if (h_model_set_expression(model, 15, expr) != 0)
    return cc;
if ((expr = h_string_parse_expression((h_string_t)s[80], model, 0, 0)) == 0)
    return cc;
if (h_model_set_expression(model, 16, expr) != 0)
    return cc;
if ((expr = h_string_parse_expression((h_string_t)s[114], model, 0, 0)) == 0)
    return cc;
if (h_model_set_expression(model, 17, expr) != 0)
    return cc;
if ((expr = h_string_parse_expression((h_string_t)s[115], model, 0, 0)) == 0)
    return cc;
if (h_model_set_expression(model, 18, expr) != 0)
    return cc;
if ((expr = h_string_parse_expression((h_string_t)s[116], model, 0, 0)) == 0)
    return cc;
if (h_model_set_expression(model, 19, expr) != 0)
    return cc;
if ((expr = h_string_parse_expression((h_string_t)s[80], model, 0, 0)) == 0)
    return cc;
if (h_model_set_expression(model, 20, expr) != 0)
    return cc;
if ((expr = h_string_parse_expression((h_string_t)s[117], model, 0, 0)) == 0)
    return cc;
if (h_model_set_expression(model, 21, expr) != 0)
    return cc;
if ((expr = h_string_parse_expression((h_string_t)s[118], model, 0, 0)) == 0)
    return cc;
if (h_model_set_expression(model, 22, expr) != 0)
    return cc;
if ((expr = h_string_parse_expression((h_string_t)s[119], model, 0, 0)) == 0)
    return cc;
if (h_model_set_expression(model, 23, expr) != 0)
    return cc;
if ((expr = h_string_parse_expression((h_string_t)s[80], model, 0, 0)) == 0)
    return cc;
if (h_model_set_expression(model, 24, expr) != 0)
    return cc;
if ((expr = h_string_parse_expression((h_string_t)s[80], model, 0, 0)) == 0)
    return cc;
if (h_model_set_expression(model, 25, expr) != 0)
    return cc;
if ((expr = h_string_parse_expression((h_string_t)s[80], model, 0, 0)) == 0)
    return cc;
if (h_model_set_expression(model, 26, expr) != 0)
    return cc;
if ((expr = h_string_parse_expression((h_string_t)s[80], model, 0, 0)) == 0)
    return cc;
if (h_model_set_expression(model, 27, expr) != 0)
    return cc;
if ((expr = h_string_parse_expression((h_string_t)s[80], model, 0, 0)) == 0)
    return cc;
if (h_model_set_expression(model, 28, expr) != 0)
    return cc;
if ((expr = h_string_parse_expression((h_string_t)s[80], model, 0, 0)) == 0)
    return cc;
if (h_model_set_expression(model, 29, expr) != 0)
    return cc;
if ((expr = h_string_parse_expression((h_string_t)s[80], model, 0, 0)) == 0)
    return cc;
if (h_model_set_expression(model, 30, expr) != 0)
    return cc;
if ((expr = h_string_parse_expression((h_string_t)s[80], model, 0, 0)) == 0)
    return cc;
if (h_model_set_expression(model, 31, expr) != 0)
    return cc;
if ((expr = h_string_parse_expression((h_string_t)s[80], model, 0, 0)) == 0)
    return cc;
if (h_model_set_expression(model, 32, expr) != 0)
    return cc;
if ((expr = h_string_parse_expression((h_string_t)s[80], model, 0, 0)) == 0)
```

```
        return cc;
    if (h_model_set_expression(model, 33, expr) != 0)
        return cc;
    if ((expr = h_string_parse_expression((h_string_t)s[80], model, 0, 0)) == 0)
        return cc;
    if (h_model_set_expression(model, 34, expr) != 0)
        return cc;
    if ((expr = h_string_parse_expression((h_string_t)s[80], model, 0, 0)) == 0)
        return cc;
    if (h_model_set_expression(model, 35, expr) != 0)
        return cc;

    if ((t = h_node_get_table(node_COMPLETE_DST_manual_C3)) == 0 || (d = h_table_get_data(t)) == 0)
        return cc;
    {
        static h_number_t data[] = {1.0,0.0,1.0};
        memcpy(d, arr_0_0, sizeof(h_number_t)*13);
        memcpy(d+14, arr_0_0, sizeof(h_number_t)*9);
        memcpy(d+24, arr_0_0, sizeof(h_number_t)*15);
        memcpy(d+40, arr_0_0, sizeof(h_number_t)*3);
        memcpy(d+44, arr_0_0, sizeof(h_number_t)*24);
        memcpy(d+68, data, sizeof(h_number_t)*3);
        memcpy(d+71, arr_0_0, sizeof(h_number_t)*19);
        memcpy(d+91, arr_0_0, sizeof(h_number_t)*11);
        memcpy(d+103, arr_0_0, sizeof(h_number_t)*15);
        memcpy(d+119, arr_0_0, sizeof(h_number_t)*8);
        memcpy(d+128, arr_0_0, sizeof(h_number_t)*14);
        memcpy(d+143, arr_0_0, sizeof(h_number_t)*16);
        memcpy(d+160, arr_0_0, sizeof(h_number_t)*10);
        memcpy(d+171, arr_0_0, sizeof(h_number_t)*13);
        memcpy(d+185, arr_0_0, sizeof(h_number_t)*18);
        memcpy(d+204, arr_0_0, sizeof(h_number_t)*9);
        memcpy(d+214, arr_0_0, sizeof(h_number_t)*16);
        memcpy(d+231, arr_0_0, sizeof(h_number_t)*11);
        memcpy(d+243, arr_0_0, sizeof(h_number_t)*19);
        memcpy(d+263, arr_0_0, sizeof(h_number_t)*11);
        memcpy(d+275, arr_0_0, sizeof(h_number_t)*5);
    }

    if ((t = h_node_get_table(node_COMPLETE_DST_manual_C8)) == 0 || (d = h_table_get_data(t)) == 0)
        return cc;
    {
        static h_number_t data[] = {1.58873E-4,0.00301329,0.34364,0.403583,0.248493,0.555222,0.370168,0.0740806,0.0106709,0.0319514,0.916743,0.039106,2.14085E-4,3.97537E-
4,2.14085E-4,3.06324E-5,3.06324E-5,2.14085E-4,0.112046,0.00610589,0.677391,0.184323,0.0189771,5.09397E-4,0.00286012,0.442681,0.5534,2.74324E-
4,0.0560839,0.283236,0.51303,0.131625,0.0154078};
        memcpy(d, data, sizeof(h_number_t)*5);
        memcpy(d+5, arr_1_58873E__4, sizeof(h_number_t)*7);
        memcpy(d+12, arr_5_88568E__5, sizeof(h_number_t)*2);
        memcpy(d+14, data+5, sizeof(h_number_t)*3);
        memcpy(d+17, arr_5_88568E__5, sizeof(h_number_t)*7);
        memcpy(d+24, data+8, sizeof(h_number_t)*8);
        memcpy(d+32, arr_2_14085E__4, sizeof(h_number_t)*2);
        memcpy(d+34, data+16, sizeof(h_number_t)*7);
        memcpy(d+41, arr_1_65322E__4, sizeof(h_number_t)*7);
        memcpy(d+48, data+23, sizeof(h_number_t)*5);
        memcpy(d+53, arr_3_92517E__5, sizeof(h_number_t)*7);
        memcpy(d+60, data+28, sizeof(h_number_t)*5);
        memcpy(d+65, arr_8_82073E__5, sizeof(h_number_t)*7);
    }

    if ((t = h_node_get_table(node_COMPLETE_DST_manual_C4)) == 0 || (d = h_table_get_data(t)) == 0)
        return cc;
    {
        static h_number_t data[] = {0.0,0.0,1.0,0.0,1.0,1.0,0.0,1.0,0.0,0.0,1.0,0.0,1.0,0.0,1.0,1.0,0.0,1.0,0.0,1.0};
        memcpy(d, data, sizeof(h_number_t));
        memcpy(d+3, data+1, sizeof(h_number_t)*4);
        memcpy(d+7, arr_0_0, sizeof(h_number_t)*2);
        memcpy(d+11, arr_0_0, sizeof(h_number_t)*2);
        memcpy(d+13, data+5, sizeof(h_number_t)*2);
        memcpy(d+17, arr_0_0, sizeof(h_number_t)*2);
        memcpy(d+19, data+7, sizeof(h_number_t)*2);
        memcpy(d+23, data+9, sizeof(h_number_t)*6);
        memcpy(d+29, arr_0_0, sizeof(h_number_t)*2);
        memcpy(d+33, arr_0_0, sizeof(h_number_t)*2);
        memcpy(d+35, data+15, sizeof(h_number_t)*5);
    }

    if ((t = h_node_get_table(node_COMPLETE_DST_manual_U4)) == 0 || (d = h_table_get_data(t)) == 0)
        return cc;
    {
        static h_number_t data[] = {-2500.0,-3500.0,-4500.0,-5500.0,-7000.0,-9000.0,-11000.0,-13000.0,-15000.0,-17000.0,-19000.0,-21000.0,-23000.0};
        memcpy(d, data, sizeof(h_number_t)*13);
```

```
}

node_buffer[0] = node_COMPLETE_DST_manual_D6;
node_buffer[1] = 0;
if ((model = h_node_new_model(node_COMPLETE_DST_manual_C20, node_buffer)) == 0)
    return cc;
if ((expr = h_string_parse_expression((h_string_t)s[120], model, 0, 0)) == 0)
    return cc;
if (h_model_set_expression(model, 0, expr) != 0)
    return cc;
if ((expr = h_string_parse_expression((h_string_t)s[121], model, 0, 0)) == 0)
    return cc;
if (h_model_set_expression(model, 1, expr) != 0)
    return cc;

node_buffer[0] = node_COMPLETE_DST_manual_C1;
node_buffer[1] = 0;
if ((model = h_node_new_model(node_COMPLETE_DST_manual_C11, node_buffer)) == 0)
    return cc;
if ((expr = h_string_parse_expression((h_string_t)s[122], model, 0, 0)) == 0)
    return cc;
if (h_model_set_expression(model, 0, expr) != 0)
    return cc;
if ((expr = h_string_parse_expression((h_string_t)s[123], model, 0, 0)) == 0)
    return cc;
if (h_model_set_expression(model, 1, expr) != 0)
    return cc;
if ((expr = h_string_parse_expression((h_string_t)s[124], model, 0, 0)) == 0)
    return cc;
if (h_model_set_expression(model, 2, expr) != 0)
    return cc;
if ((expr = h_string_parse_expression((h_string_t)s[125], model, 0, 0)) == 0)
    return cc;
if (h_model_set_expression(model, 3, expr) != 0)
    return cc;
if ((expr = h_string_parse_expression((h_string_t)s[126], model, 0, 0)) == 0)
    return cc;
if (h_model_set_expression(model, 4, expr) != 0)
    return cc;
if ((expr = h_string_parse_expression((h_string_t)s[127], model, 0, 0)) == 0)
    return cc;
if (h_model_set_expression(model, 5, expr) != 0)
    return cc;

node_buffer[0] = node_COMPLETE_DST_manual_C18;
node_buffer[1] = node_COMPLETE_DST_manual_C6;
node_buffer[2] = 0;
if ((model = h_node_new_model(node_COMPLETE_DST_manual_U3, node_buffer)) == 0)
    return cc;
if ((expr = h_string_parse_expression((h_string_t)s[128], model, 0, 0)) == 0)
    return cc;
if (h_model_set_expression(model, 0, expr) != 0)
    return cc;
if ((expr = h_string_parse_expression((h_string_t)s[129], model, 0, 0)) == 0)
    return cc;
if (h_model_set_expression(model, 1, expr) != 0)
    return cc;
if ((expr = h_string_parse_expression((h_string_t)s[130], model, 0, 0)) == 0)
    return cc;
if (h_model_set_expression(model, 2, expr) != 0)
    return cc;
if ((expr = h_string_parse_expression((h_string_t)s[131], model, 0, 0)) == 0)
    return cc;
if (h_model_set_expression(model, 3, expr) != 0)
    return cc;
if ((expr = h_string_parse_expression((h_string_t)s[132], model, 0, 0)) == 0)
    return cc;
if (h_model_set_expression(model, 4, expr) != 0)
    return cc;
if ((expr = h_string_parse_expression((h_string_t)s[130], model, 0, 0)) == 0)
    return cc;
if (h_model_set_expression(model, 5, expr) != 0)
    return cc;

if ((t = h_node_get_table(node_COMPLETE_DST_manual_U2)) == 0 || (d = h_table_get_data(t)) == 0)
    return cc;
{
    static h_number_t data[] = {0.0,-12.5,-37.5,-75.0,-300.0,-750.0,-3000.0,-7500.0,-12500.0,-17500.0,-22500.0,-27500.0,-32500.0,-37500.0,-42500.0};
    memcpy(d, data, sizeof(h_number_t)*15);
}

node_buffer[0] = node_COMPLETE_DST_manual_D6;
```

```
    node_buffer[1] = 0;
    if ((model = h_node_new_model(node_COMPLETE_DST_manual_U1, node_buffer)) == 0)
        return cc;
    if ((expr = h_string_parse_expression((h_string_t)s[133], model, 0, 0)) == 0)
        return cc;
    if (h_model_set_expression(model, 0, expr) != 0)
        return cc;
    if ((expr = h_string_parse_expression((h_string_t)s[134], model, 0, 0)) == 0)
        return cc;
    if (h_model_set_expression(model, 1, expr) != 0)
        return cc;

    return cc;
}
```